\documentclass[rmp,aps,graphicx,floatfix,amssymb,twocolumn]{revtex4}
\usepackage{amsmath}
\usepackage{amstext}
\usepackage{latexsym}
\usepackage{graphicx}
\usepackage{bm}
\usepackage[colorlinks=true, allcolors=blue]{hyperref}
\usepackage{color}

\usepackage[disable]{todonotes} 
 \usepackage[final]{changes} % Implement all changes without leaving comment 

\newcommand{\note}[1][pink]{\todo[inline, color=#1]}

\def\be{\begin{eqnarray}}
\def\bea{\begin{eqnarray}}
\def\bma{\begin{mathletters}}
\def\ee{\end{eqnarray}}
\def\eea{\end{eqnarray}}
\def\ema{\end{mathletters}}

\newcommand{\nc}{\newcommand}
\nc{\rnc}{\renewcommand} 
\nc{\proj}[1]{\ket{#1}\bra{#1}} \rnc{\vec}{\mathbf}
\nc{\braket}[2]{\langle\, #1\,|\,#2\,\rangle}
\nc{\half}{\frac{1}{2}}
\nc{\vfigure}[3]{
\begin{figure}[th]
\centerline{\psfig{file=figures/#1.eps,width=#2}}
\vspace*{8pt}
\caption{#3}
\end{figure}}
\nc{\vpstexfigure}[3]{
\vfigure{#1}{#2}{#3}}
\nc{\prj}{\mathcal{P}} \nc{\hilb}{\mathcal{H}}
\nc{\pth}{\mathcal{C}} \nc{\inprod}[2]{\braket{#1}{#2}}
\nc{\upket}{\ket{\uparrow}} \nc{\downket}{\ket{\downarrow}}
\def\CC{{\rm\kern.24em \vrule width.04em height1.46ex depth-.07ex
\kern-.30em C}}
\def\P{{\rm I\kern-.25em P}}
\def\N{{\rm I\kern-.25em N}}
\def\RR{{\rm
 \vrule width.04em height1.58ex depth-.0ex
 \kern-.04em R}}
\def\id{{\rm 1\kern-.26em l}}
\def\ZZ{{\sf Z\kern-.44em Z}}

\def\i{{\rm i}\,}

\newenvironment{eqblock}[2]{\beq\label{#2}\begin{array}{#1}}{\end{array}
  \eeq}
\newenvironment{neqblock}[1]{\[\begin{array}{#1}}{\end{array}\]}

\newcommand{\beqb}{\begin{eqblock}}
\newcommand{\eeqb}{\end{eqblock}} 
\newcommand{\nbeqb}{\begin{neqblock}}
\newcommand{\neeqb}{\end{neqblock}} 

\newcommand{\beq}{\begin{equation}}
\newcommand{\beqa}{\begin{eqnarray}}
\newcommand{\eeq}{\end{equation}}
\newcommand{\eeqa}{\end{eqnarray}}
\newcommand{\nbeqa}{\begin{eqnarray*}}
\newcommand{\neeqa}{\end{eqnarray*}}

\graphicspath{{./figures/},{./figuresAppendix/}}

\newcommand{\cn}[2]{\ensuremath{{#1}^\dagger_{#2}}}
\newcommand{\nn}[2]{\ensuremath{{n}^{#1}_{#2}}}

\newcommand{\vc}[1]{\ensuremath{{\bf#1}}}

\newcommand{\avg}[1]{\ensuremath{\langle#1\rangle}}

\newcommand{\op}[1]{\ensuremath{\hat{#1}}}

\begin{document}
\note{WK: Note that we currently have 24 pages and need to cut down to 20p}
\listoftodos
\clearpage

\title{Atomtronic circuits: from many-body physics to  quantum technologies}

\author{Luigi Amico}
\affiliation{Quantum Research Centre, Technology Innovation Institute, Abu Dhabi, UAE}
\altaffiliation{On leave from Dipartimento di Fisica e Astronomia, Via S. Sofia 64, 95127 Catania, Italy}
\affiliation{INFN-Sezione di Catania, Via S. Sofia 64, 95127 Catania, Italy}
\affiliation{Centre for Quantum Technologies, National University of Singapore, 3 Science Drive 2, Singapore 117543, Singapore}
\affiliation{LANEF {\it 'Chaire d'excellence'}, Universit\`e Grenoble-Alpes $\&$ CNRS, F-38000 Grenoble, France}
\author{Dana Anderson}
\affiliation{Department of Physics and JILA, University of Colorado, Boulder, Colorado, 80309-0440, USA}
\author{Malcolm Boshier}
\affiliation{MPA Division, Los Alamos National Laboratory, Los Alamos, NM 87545, USA}
\author{Jean-Philippe Brantut}
\affiliation{Institute of Physics, EPFL, 1015 Lausanne, Switzerland}
\author{Leong-Chuan Kwek}
\affiliation{Centre for Quantum Technologies, National University of Singapore,
	3 Science Drive 2, Singapore 117543, Singapore}
\affiliation{MajuLab, CNRS-UNS-NUS-NTU International Joint Research Unit, UMI 3654, Singapore}
\affiliation{Institute of Advanced Studies, Nanyang Technological University,
	60 Nanyang View, Singapore 639673, Singapore}
\author{Anna Minguzzi}
\affiliation{Universit\'{e} Grenoble-Alpes and CNRS, LPMMC, F-38000 Grenoble}
\author{Wolf von Klitzing}
\affiliation{Institute of Electronic Structure and Laser, Foundation for Research and Technology-Hellas, Crete, Heraklion 70013, Greece}

%\pacs{}

\begin{abstract}
Atomtronics is an emerging field that aims to manipulate ultracold atom moving in matter wave circuits for both fundamental studies in quantum science and technological applications. In this colloquium, we review recent progress in matter-wave circuitry and atomtronics-based quantum technology. After a short introduction to the basic physical principles and the key experimental techniques needed to realize atomtronic systems, we describe the physics of matter-waves in simple circuits such as ring traps and two-terminal systems. The main experimental observations and outstanding questions are discussed. We also present possible applications to a broad range of quantum technologies, from quantum sensing with atom interferometry to future quantum simulation and quantum computation architectures. 
\end{abstract}

\maketitle

%\newpage \clearpage \newpage

\tableofcontents

%\newpage \clearpage \newpage
\section{Introduction} 
\label{intro}

Atomtronics is the emerging quantum technology of matter-wave circuits which coherently guide propagating ultra-cold atoms  \cite{seaman2007atomtronics,amico2005quantum,amico2017focus,amico2021roadmap}.   Developing and applying such circuits has been a goal of cold atom physics for decades: see, for example, the opening paragraphs in \onlinecite{muller1999guiding, dekker2000guiding, schneble2003integrated, leanhardt2002propagation,dumke2002interferometer}.  Realizing this vision was an important motivation for the invention and development of atom chip technology in the early years of this century  \cite{schmiedmayer1995wire,schmiedmayer1996classical,schmiedmayer1996particle,denschlag1999neutral,denschlag1999guiding}. 
While this approach has not yet demonstrated coherent propagation of guided
matter waves, there is an extensive body of work on {coherent manipulation of
trapped clouds} of ultracold atoms on atom chips which provides a foundation
for the more recent work discussed here.
%While this approach has not yet demonstrated coherent propagation of guided matter waves, there is an extensive body of work with static clouds of %ultracold atoms trapped on atom chips which provides a foundation for the more recent work discussed here. 
This research is reviewed in  \cite{folman2002microscopic, reichel2002microchip,fortagh2007magnetic,reichel2011atom,keil2016fifteen}.  In the implementations of atomtronic circuits realized to date, matter waves travel in guides made of laser light or magnetic fields. 
These approaches offer highly controllable, flexible and versatile platforms at the microscopic spatial scale \cite{henderson2009experimental, rubinsztein-dunlop2016roadmap,Gauthier:2019aa}. 
The quantum fluid flowing through atomtronic circuits is provided by ultracold atoms that can be fermions, bosons, or a mixture of the two species. 
Cold atom quantum technology allows coherent matter-wave manipulations with unprecedented control and precision over a wide range of spatial lengths and physical conditions. \cite{ketterle2002nobel,cornell2002nobel,bloch2005ultracold,dalfovo1999theory}. 

Atomtronic circuits are suitable as
%define
cold-atom quantum simulators \cite{lewenstein2012ultracold, dowling2003quantum,bloch2005ultracold,buluta2009quantum,cirac2012goals,lamata2014epj} in which matter wave currents are harnessed as probes to explore the physics of the system. 
%Atomtronic circuits grant access to both basic and applied research and technological applications. 
In this way, important problems in fundamental quantum science, such as
superfluidity, strong correlations in extended systems, 
topological aspects in quantum matter,
quantum transport, and various mesoscopic effects, can be studied from a new angle \cite{stadler2012observing,husmann2015connecting,valtolina2015josephson,krinner2016mapping,burchianti2018connecting,Del-Pace:2021ti}.

At the same time, atomtronic circuits play an important role in applied science and technology.
Like electronic devices, atomtronic circuits operate over a separation of time and length scales between devices and leads. 
This permits the construction of standardized functional units connected to each other by waveguides acting as wires.
Atomtronic counterparts of known electronic or quantum electronic components have been the first developments in the field. Some examples include atomtronic amplifiers, diodes, switches, batteries, and memories \cite{seaman2007atomtronics,pepino2009atomtronic,pepino2021advances,anderson2021matterwaves,caliga2017experimental,zozulya2013principles,stickney2007transistorlike,caliga2016principles}. Moreover, cold atom realizations of Josephson junctions have led to the fabrication and analysis of atomtronic superconducting quantum interference devices (SQUIDs) \cite{eckel2014hysteresis,jendrzejewski2014resistive,wright2013Driving,ramanathan2011superflow,ryu2013experimental,amico2014superfluid,aghamalyan2015coherent,haug2018readout, ryu2020quantum}. 
Atomtronics can also contribute to the field of quantum sensors \cite{degen2017quantum,cronin2009optics,bongs2019taking}. Building on the pioneering demonstrations of compact atom interferometers using {trapped} Bose-Einstein condensates (BECs) \cite{schumm2005matter,gunther2007atom,jo2007phase,bohi2009coherent, riedel2010atom}, several solutions for compact atomtronic interferometers with enhanced sensitivity to inertial forces and electromagnetic fields have been studied \cite{wang2005atom,akatsuka2017optically,ryu2015integrated,burke2009scalable,moan2020quantum,qi2017magnetically,wu2007demonstration,krzyzanowska2022matter,kim2022one,mcdonald2014bright,mcdonald2013optically}.
 %
%Finally, we note that highly sophisticated devices and enhanced functionalities with no analog in electronics nor photonics can also be designed and built %from microscopic descriptions with the aforementioned enhanced control \cite{lewenstein2012ultracold,bloch2008many}.
%
{Finally, we observe that the aforementioned specific properties of coherence, control and flexibility characterizing ultracold matter-wave circuits can enable devices with no direct analog in electronics or photonics technology. Proofs of concept built on features inherent to specified microscopic implementations and combined with  specifically suited enabling technologies have  been considered recently\cite{lau2022atomtronic,naldesi2022enhancing,chetcuti2022persistent,amico2014superfluid,aghamalyan2013effective,krzyzanowska2022matter,kim2022one}}.

{In this Colloquium, we provide a short and accessible review of the atomtronics field  for a broad educated audience of researchers. 
For a more technical  discussions of some of the most recent developments, we refer the reader to the roadmap article \onlinecite{amico2021roadmap}}.
%Further details and a more comprehensive discussion of  specific aspects of the field can be found in in the roadmap article\onlinecite{amico2021roadmap}}.
The Colloquium is organized as follows: In Sec.~\ref{PhysPrinc}, we discuss the state of the art in optical and magnetic trapping technologies that lead to a variety of circuits. In Sec.~\ref{CoherentMesoscopics}, we focus on the coherent flow in simple atomtronic networks of mesoscopic size. In this section, we bridge many-body models with persistent currents and two-terminal transport through a mesoscopic channel. In Sec.~\ref{components} we describe some of the components that have been studied and developed so far. Finally, we conclude and provide an outlook in Sec.\ref{future}. 

\section{Traps and Guides} 
\label{PhysPrinc}
Atomotronics has been  made possible by the ability to trap matter waves of coherent cold atoms in  complex smooth potentials in which matter waves can be  feasibily created, guided, and manipulated in controllable and flexible fashion. These potentials are produced either by optical fields that exert forces on atoms through their polarizability or by magnetic fields that create forces on atomic magnetic dipoles. 

\subsection{Optical Potentials}
\label{optical}
%B. Optical dipole potentials for atomtronics

% \subsubsection{Introduction}

%Optical potentials formed by static or dynamic laser beams are a particularly appealing technology for realizing atomtronic circuitry. 
The formation of optical potentials through static or dynamic laser beams is a mature technology for the realization of atomtronic circuitry.
The flexible potentials can have almost arbitrary complexity in both space and time domains.

Optical manipulation of ultracold atoms is based on the electric dipole interaction between the atoms and the laser beam. When the laser frequency $\omega$ is far-detuned from an atomic transition of frequency $\omega_0$, the interaction energy takes the form of an optical dipole potential
$ %\begin{eqnarray}
\displaystyle{U(\mathbf{r}) =- \frac{3  \pi c^2}{2 \omega_0^3} \frac{\Gamma}{\Delta} I(\mathbf{r})}, 
$ %\end{eqnarray}
where $\Delta=\omega - \omega_0$ is the detuning, $\Gamma$ is the natural decay rate of the population of the excited state, and $I(\mathbf{r})$ is the position-dependent laser intensity. This dipole force can be attractive ($\Delta <0$ or ``red-detuned") or repulsive ($\Delta > 0$ or ``blue-detuned") \cite{grimm2000optical}. 
The detuning should be large enough so that spontaneous scattering is negligible on the timescale of the experiment.

\subsubsection{Static laser beams}
Waveguides supporting coherent propagation of matter waves must be smooth to avoid excitations out the guide ground state to higher modes, and stable because fluctuations in the potential cause fluctuations in the phase accumulated by the matter wave. A collimated laser beam is a straightforward solution to this problem. The first guides for cold, non-condensed atoms used the evanescent field of blue-detuned light propagating in a hollow optical waveguide \cite{renn1996evanescent, muller2000guiding, rhodes2002guiding}. This is followed by the introduction of traps and guides based on hollow blue-detuned laser beams created with doughnut or Laguerre-Gaussian transverse modes that removed the need for a material optical guide \cite{kuga1997novel}. This approach enabled creation of the first waveguide for a Bose-Einstein condensate (BEC) \cite{bongs2001waveguide}. When the Laguerre-Gaussian beam is tightly focused, the optical dipole potential becomes more like a toroidal trap \cite{olson2007cold} than a waveguide. Waveguide potentials can also be realized with Bessel beams \cite{arlt2000atom}. 

Red-detuned collimated laser beam is a simpler technology for creating atomtronic waveguides when spontaneous emission is sufficiently small. An early example of this approach is the realization of a simple beamsplitter for propagating cold thermal atoms with a pair of crossed red-detuned laser beams \cite{houde2000cold}. Subsequent demonstrations include coherent propagation of BEC matter wavepackets to realize a Mach-Zehnder atom interferometer \cite{mcdonald2013optically, kim2022one}, a beamsplitter for BECs \cite{gattobigio2012optically} and a waveguide Sagnac atom interferometer \cite{krzyzanowska2022matter}. Red-detuned collimated lasers are also used to guide the matter wave produced by an atom laser \cite{Guerin:2006aa,Couvert:2008mh,dall2010transverse}. A very recent development is the use of clipped gaussian beams to create elongated trapping and guiding potentials \cite{lim2021large}.

%\subsubsection{Other approaches}
Besides the standard approaches mentioned above,  microfabricated optical elements  \cite{birkl2001atom}, arrays of micro-lenses  %used to create potential geometries analogous to a large array of Mach-Zehnder interferometers 
\cite{dumke2002interferometer} and the application of conical refraction in a biaxial crystal  \cite{turpin2015blue} have been proposed as sophisticated routes to realize complex circuits.
Standing waves of laser light 
%form periodic potentials that are used in atomtronic circuitry. For example, standing waves 
impressed on a collimated laser waveguide have been shown to form a distributed Bragg reflector \cite{fabre2011realization}. Pulsed optical standing waves are also employed as beam splitters \cite{wang2005atom, wu2005splitting, kim2022one, krzyzanowska2022matter}.

\subsubsection{Time-averaged optical potentials}
\label{timeaveragedpotentials}
\begin{figure}
\centering
\hspace*{-0.5cm}\includegraphics[width=0.52\textwidth]{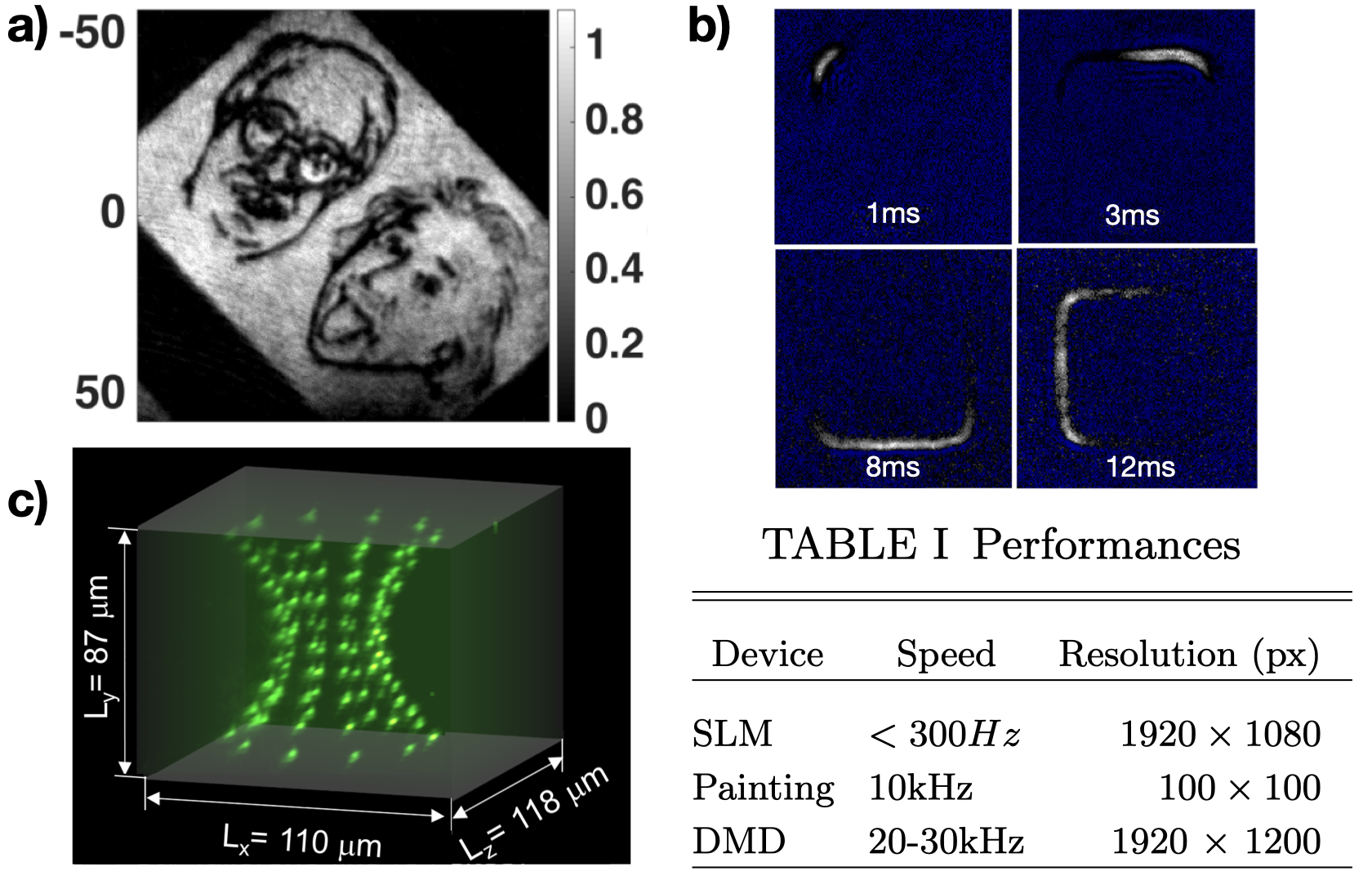}
 \caption{A Bose-Einstein condensate trapped  in potential made by a)  DMD, lenght scales in $\mu$m, b) painting technique, dimension of each image is $70 \times 70 \mu$m, c) SLM. The table compares the performances of the devices in terms of refreshing time, resolution, and diffraction efficiency. Figures a) adapted from  \onlinecite{gauthier2016direct}; b) adapted from \onlinecite{ryu2015integrated},c) adapted from \onlinecite{barredo2018synthetic}.
% \note{Luigi: Permission needed for a and c}
 }
\label{fig:sculpted}
\end{figure}

Optical dipole potentials based on static laser beams are cylindrically symmetric and they can have no time dependence beyond a scaling of the trap strength.
This shortcoming motivated the development of time-averaged optical potentials.
Similar to the guides discussed above, the initial experiments with this approach used non-condensed thermal atoms, confining them to box and stadium potentials with walls formed by a blue-detuned laser beams that is rapidly scanned with a pair of acousto-optic deflectors \cite{friedman2001observation, milner2001optical}. While early experiments on trapping Bose-Einstein condensates (BECs) in multiple wells, formed by rapidly switching the position of a single red-detuned laser, find that the condensates are heated \cite{onofrio2000surface}, that issue is absent in later work in which a time-averaged tightly-focused laser beam "paints" a desired potential on a canvas provided by a light sheet that confined atoms to a horizontal plane. 
This ``painted potential" \cite{henderson2009experimental, schnelle2008versatile} is able to realize arbitrary and dynamic 2D matter waveguide structures-see Fig.\ref{fig:sculpted}a).  This includes the important case of toroidal potentials, \cite{henderson2009experimental, ryu2014creation, bell2016bose}, where periodically reducing the intensity the laser painting the attractive potential creates movable repulsive barriers that can form Josephson junctions in an atom SQUID geometry \cite{ryu2013experimental, ryu2020quantum}. Repulsive barriers can also be imposed on a trap by painting with a blue-detuned laser \cite{wright2013Driving, ramanathan2011superflow}.
A significant advantage of this approach is that a suitable modulation of the tweezer beam intensity as it paints the atomtronic circuit can flatten out any imperfections in the potential \cite{ryu2015integrated, bell2016bose}, enabling creation of waveguides smooth enough to support single mode propagation and to realize the first coherent beam splitter for propagating matter waves \cite{ryu2015integrated}.

While painting has the advantages of making efficient use of laser power and enabling fine control of the shape of the potential landscape, it has some limitations. The time averaging requirement that the potential be painted at a rate significantly higher than the guide trapping frequency limits the trapping frequencies attainable with current acousto-optic deflector technology to several kHz. While it is usually a good approximation to regard the painted potential as static, in some circumstances the time-varying phase imprinted by the painting beam can be an issue \cite{bell2018phase}.

\subsubsection{Spatial light modulators and Digital Micro-Mirror Devices}
A second technology for creating complex 2D potential landscapes on a light sheet relies on spatial light modulators (SLMs) that can impose amplitude or phase modulation on a laser beam which forms the desired potential after propagation through suitable optics. Two approaches have been demonstrated: a Fourier optics approach in which the SLM acts as a hologram, and a direct imaging of an intensity pattern formed by the SLM. A detailed discussion of the production of arbitrary optical potentials is presented in reference \cite{gauthier2021Dynamic}. A recent demonstration for SLM realization of 3d-potentials was reported in \onlinecite{barredo2018synthetic}c).

Early work in this direction used liquid crystal modulators to create phase holograms producing arrays of tweezer beams \cite{curtis2002Dynamic} or more complex geometries \cite{boyer2004dynamic, boyer2006dynamic, gaunt2012robust}. Liquid crystal SLMs are now widely used in creating dynamic optical tweezer systems for assembling arrays of Rydberg atoms used for quantum information processing \cite{Nogrette:2014aa}. Advantages of liquid crystal modulators include the ability to impose phase or amplitude modulation on an optical beam and the possibility of using them either as holographic elements or for direct imaging, as well as a high power efficiency. The disadvantages include a limited response time for creating time-dependent potentials, as well as the technical overhead of computing real-time SLM holograms for dynamic potentials, which can be addressed using increasingly powerful GPUs. 

%flicker modulation of the potential resulting from the periodic refresh of the pixels and limited response time for creating time-dependent potential \cite{Hueck:2017tv}. 

An alternative to SLMs are digital micromirror devices (DMDs), producing binary patterns over a matrix of individually switched mirrors. The intensity pattern formed by the DMD can be imaged directly onto an atomic cloud using standard imaging techniques to form intricate potentials in the image plane of the optical system. Fine intensity control overcoming the binary nature of the DMD can then be achieved through half-toning techniques \cite{gauthier2016direct, gauthier2021Dynamic, tajik2019designing,kumar2016minimally,zou2021optical}. DMDs can also be used as programmable diffraction gratings, similar to SLMs, at the expenses of low power efficiency \cite{zupancic2016ultra}. However, they offer higher refresh rate, allowing for their use in dynamical experiments \cite{Ha:2015vu}. Fig.~\ref{fig:sculpted}b)  illustrates the power of this technique for creating a BEC in the shape of a sketch of Bose and Einstein. DMDs do suffer from flicker modulation due to their intended use in image projectors for the consumer market and so some devices may require customization for optimal use \cite{Hueck:2017tv}.

The performances of painting techniques, SLM, and DMD are summarized in the table of Fig. \ref{fig:sculpted}. The three thecniques share a diffraction efficiency of $\sim 65 - 80 \%$.
 
\subsection{Magnetic potentials}
\subsubsection{Magnetic traps}
%%%%%%%%%%%%%%%%%%%%%%%%%%%%%%%%%%%%%%%
% \JeanPhilippen{I suggest to remove the detailed expressions for the basic shapes of magnetic, unless they are actually needed for later descriptions.}
Magnetic trapping lies at the heart of many cold atoms quantum technologies. %sorry there is no such review... \cite{General reviews on magnetic trapping} \Wolfn{I just cannot find a review on this... anyone?}. 
Here, we sketch the logic of the technique.
%Possibly the simplest atomtronic device is the
Magnetic traps confine spin-polarized atoms of non-zero magnetic moment to a local minimum of a static magnetic field $\vec{B}$.
If the magnetic field at the center of mass position of an atom is sufficiently large and varies slowly, then its spin follows its change in direction and magnitude.
The Zeeman energy of spin polarized atoms $(V =-\bm{\mu} \cdot \vec{B})$ can then be written as $V=m_\mathrm{F} g_\mathrm{F} \mu_\mathrm{B} \left| \vec{B} \right|$, 
with $ m_\mathrm{F} = \{ -F \ldots F\} $ being the magnetic hyperfine number, $g_\mathrm{F}$ the Land\'e $g$-factor, and $\mu_\mathrm{B}$ the Bohr magneton. 
Unfortunately, Maxwell's equations forbid the generation of a dc-magnetic field maximum in free space. 
Therefore, one has to trap so-called low-field seeking states, whose energy increases with magnetic field strength. 
The field strength of the magnetic minimum has to be sufficiently large in order to prevent non adiabatic spin-flip transitions to lower-energy (high-field seeking) states. The latter can cause atoms to be expelled from the trap \cite{majorana1932atomi}.
The two most common magnetic configurations are the Ioffe-Pritchard (IP) \cite{baiborodov1963adiabatic, Pritchard1983Cooling} and the time-orbiting potential (TOP) \cite{petrich1995stable} traps.

An IP-trap consists of a radial quadrupole and an axial parabolic field, which together generate an elongated local minimum in the magnetic field. 
% for a very elongated trap is:
% \begin{align}
%  \label{eq:B-IP}
%  \vec{B}_\mathrm{IP} &= \alpha \left( x \, \hat{\bm{e}}_\mathrm{x}
%  - y \, \hat{\mathbf{e}}_\mathrm{y}\right)
%  +\left(B_0 +\frac{1}{2}\beta\,z^2 \right)\, \hat{\mathbf{e}}_\mathrm{z}
% % \\ V_\mathrm{IP}&=\sqrt{\alpha x^2+\alpha y^2+ },
% \end{align}
% where $\alpha$ is the radial gradient, $\beta$ the curvature of the trap, $B_0$ the offset field in the center of the trap and $\hat{\bm{e}}_\mathrm{x}$, $\hat{\bm{e}}_\mathrm{y}$, and $\hat{\bm{e}}_\mathrm{z}$ are the normal vectors of the coordinate system.
% The strength of the confinement is described by the radial and axial trapping frequencies, which are $\omega_\rho=\left(m_\mathrm{F} g_\mathrm{F} \mu_\mathrm{B} \,m^{-1}\alpha^2/B_0\right)^{1/2}$ and
%  $\omega_\mathrm{z}=\left(m_\mathrm{F} g_\mathrm{F} \mu_\mathrm{B}\,m^{-1} \beta \right)^{1/2}$ 
%  respectively, where $m$ is the mass of a trapped atom.
Typical values for the radial trap frequency 
range from few hundreds for macroscopic traps  to few thousands Hz %\Annan{Is 400 and 4000 a strict limit? I put %simeq, OK? in alternative, we could say 'few %hundreds to few thousands'} from %$\omega_\rho\simeq 2\pi \,400$\,Hz for %macroscopic traps to $\omega_\rho\simeq 2\pi %\,4000$\,Hz 
for chip-based traps \cite{hansel2001bose}. Typical axial frequencies are a few tens of hertz.
Macroscopic IP traps are usually formed from a combination of large race-track shaped coils for the radial gradient and small `pinch' coils for the parabolic axial field.
%%%%%%%%%%%%%%%%%%%%%%%%%%%%%%%%%%%%%%%
%\subsubsection{Permanent Magnets}
%%%%%%%%%%%%%%%%%%%%%%%%%%%%%%%%%%%%%%%
% A few alternatives to using external wires to manipulate magnetic fields have emerged. 
It is also possible to use structures from permanently magnetized materials, allowing the creation of larger magnetic gradients and thus steeper traps. 
They also provide a larger degree of freedom in design when compared to their purely electro-magnetic counterparts albeit at the cost of an inability to dynamically change the strength of the confinement or easily release the atoms from the trap \cite{tollett1995permanent, Davis1999JOOB,Sinclair2005PRA,Fernholz2008PRA}.
% An additional complication arises from the difficulty of releasing the atoms for time-of-flight (ToF) imaging.

The TOP-trap uses a static magnetic quadrupole field, the center of which is displaced away from the atoms using a rotating magnetic homogeneous offset field  \cite{petrich1995stable, hodby2000bose-einstein}.
This oscillating field may be modified locally using inductively coupled conducting structures \cite{pritchard2012Demonstration,sinuco-leon2014inductively}.
Care must be taken for the  offset field to rotate slow enough for the spin of the atoms to be able to follow, but fast enough for the center of mass of the atoms not to be significantly affected. 
If this condition is fulfilled, then the atoms will be trapped in the time-average of the magnetic potential.
For a static magnetic quadrupole field with an offset field rotating in the symmetry plane this will result in trap with the shape of an oblate sphere and the trapping frequencies $\omega_z=\sqrt{8}\omega_x=\sqrt{8}\omega_y$, with typical trapping frequencies from $\omega_z/2\pi = 40\,$Hz to 1\,kHz.  

\subsubsection{Atom chips \label{sec:AtomChips}}

% \Annan{avoid duplication with next paragraph}
A simple atom-trap can be produced by applying  a transverse homogeneous magnetic field to  the one produced by a single wire \cite{Schmiedmayer1995PRA}. 
By shaping the wires on a surface, 'Atom chip' traps \cite{reichel1999atomic, folman2000controlling, hansel2001bose} consisting of micro-sized current-carrying wires can be efficiently manufactured using standard semiconductor technologies. 
Such thin wires can be cooled very efficiently through the substrate and thus permit very large current densities resulting in very large magnetic gradients. Consequently,  trapping frequencies of  10\,kHz can be achieved. 
Another advantage is the ability to create in 2D complex wire structures \cite{folman2000controlling,fortagh2007magnetic, keil2016fifteen}.
Simple H-, T-, U-, Y- and Z-shaped wires can create a wide range of fields. 
For example, a magnetic IP trap can be formed with an elongated Z-shaped structure,  a  3D quadrupole trap  with a simple U-shaped wire \cite{reichel1999atomic}, and a 2D-quadrupole can be formed from three parallel wires, thus creating a matter wave waveguide along which atoms can be propagated \cite{folman2000controlling,Long2005EPJD}.
% 
% 
% \Wolfn{I have moved the magnetic part of the atom chips here.}
% 
%Aside from trapping cold atoms in optical traps, there is %another ingenious means to create trap, prepare, manipulate, %and measure cold atoms. , 
%Thi has its foundation built from atom interferometry % \cite{cronin2009optics}. 
% One great advantage of the chip traps, when compared to  macroscopic traps, is  the efficient cooling afforded by the small size and low power consumption, which results in much increased current densities in the conductors and thus in much stronger confinement. 
% Furthermore, the much reduced size of the chip-based conductors and their  proximity  to the atoms allow more complex structures to be achieved \cite{folman2000controlling,fortagh2007magnetic, keil2016fifteen}. 
% Atom chips use very small wires on a solid substrates (typically silicon and glass), thus enabling extremely large current densities (and thus magnetic gradients $\alpha$) and more complex geometries. 
% 
%caliga2016transport, dipole trap not chip.
%huet2012experimental too preliminary for a review paper
% The atom chip technology emerges from a fortunate coalescence of the mature semiconductor industry and atom optics. 
% 
Cryogenically cooled atom chips  have also allowed superconducting devices to be incorporated \cite{ nirrengarten2006realization, mukai2007persistent, hyafil2004coherence, salim2013high}.
Atom chips have thus become compact hybrid platforms to trap, prepare, manipulate, and measure cold atoms. 
They provide the route for miniaturization and interfacing different atomtronic components in more complex devices \cite{birkl2001atom,Gehr2010Cavity,salim2013high}.

Corrugation and noise currents in the conducting wires of an atom chip represent an important challenge to atom-chip based atomtronic circuits that allow a coherent flow of atoms over macroscopic distances \cite{folman2002microscopic, henkel2003fundamental}. %\paragraph{Noise in Atom-Chip Potentials}
%
%The resolution for the tailored fields is typically of the order of the distance from the field source. 
%So the construction of the atomic circuit at a distance of no more than a few micro-meters from the surface of the chip is required. 
%At this tiny scales, atom chips need to surmount some inherent challenges.
%Bringing atoms closer to the chip surface inevitably leads to stray fields and %other disruptive noises. 
These noises can emerge from diverse causes, ranging  current scattering due to unintended changes in flow of direction of the current, noises in the power supplies, or magnetic impurities \cite{leanhardt2002propagation,kraft2002anomalous,kruger2007potential,david2008magnetic} to thermal (Johnson) noise \cite{dikovsky2005reduction}.
%Such technical noises constitute the strongest source of noise in most atom chips. 
%The consequent atomic density variation results in the fragmentation of the atom %cloud into several separated components as it is brought
%close to the surface \cite{david2008magnetic}.
%The other type of noise is Johnson noise, whose origin stems from thermal %%energy. As the atoms approach the surface, these noises increase. Johnson noise %could be mitigated with lower conductance materialsthinner layers or alloys. %Surprisingly, it has been shown that superconducting surfaces possess low level %of noises \cite{kasch2010cold,muller2010programmable,dikovsky2009superconducting%, muller2010trapping}. Decoherence phenomena arising from these noises pose %significant problems for the preservation of coherence of the atoms as well as %the trapping and manipulation of the atoms. However, with continued improvement %in chip fabrication techniques, advancement in multi-layer structures and thin %film magnetic materials , one should be able to overcome some of the noises and %limitations in current designs.
\begin{figure}\label{fig:AtomChip}
 \includegraphics[width=0.85\columnwidth]{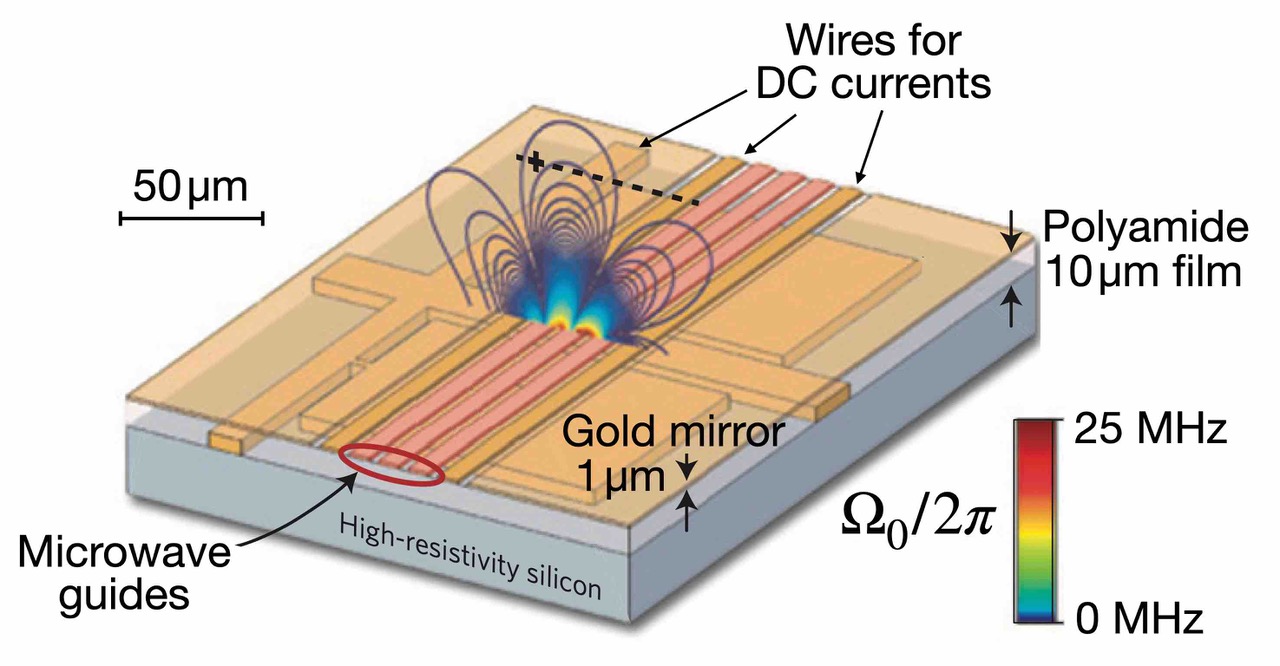} %{atomchip.png}
 \caption{A schematic drawing of an atom-chip including its magnetic potentials: Atom chips can be used for double well physics in 1D and crossover regimes, which is enabled by the ability to create robust double-well potentials using rf dressing of the atoms (Adapted from \onlinecite{bohi2009coherent}).  %\deleted[id=wk is this not a bit too specific for a caption?]{Atom chips permit fast changes of trapping parameters, which are used in optimal control schemes, allowing one to engineer the external state of the condensates, which is used, for instance, for the highly efficient creation of twin-atom beams \cite{bucker2011twin-atom}.}
% \deleted[id=WK this has little to do with atom-chips.]{Furthermore, a unique %time-of-flight fluorescence imaging system allows one to work with extremely dilute %clouds and provides single atom sensitivity, which promotes experiment in quantum %matter-wave optics, such as Hanbury Brown-Twiss type measurements of second-order %correlations in expanding clouds.}
% 
 %\Luigin{To be checked. Do we want to insert a figure here? %\Wolf{I think we should %have a figure on atom chips. I am not %sure about the caption though. Maybe we should %keep the more %general for the figure?\Luigi{How about something similar to %the chip %we used in our EU proposal?}\Wolf{That is a %drawing... right? I would prefer a photo %here. Maybe we should %ask one of the atomchip groups for a nice photo?
 % http://www.worldofindie.co.uk/?p=1801 from RAL
 % Folman: https://www.nature.com/articles/nphys426
 % space station https://www.sciencemag.org/news/2017/09/coolest-science-ever-headed-space-station
 % Rasel http://www.2physics.com/2016/12/a-compact-gravimeter-based-on-atom-chip.html 2016
 % }}}
}
 %http://atomchip.org/rubidium-ii/
% http://atomchip.org/rubidium-ii/experiment-setup/the-atom-chip/
\end{figure}
%Another type of noise related to the propagation along a waveguide on the chip %occurs when there is any corrugation of the waveguide potential. The corrugation %couples the forward motion to the transverse vibrationally excited modes causing %heating and loss of contrast in atom interferometry.
%It appears that these corrugations are almost unavoidable in magnetic fields due %to the meandering nature of the flow of electrons even if the conductor itself %is perfectly smooth.
%The corrugation of the magnetic potential drops off exponentially with distance %according to $\exp(-k \rho)/\sqrt{k\rho}$ % MINUS sign added
%\label{sec:roughness}, where $\rho$ is the distance from the conductor and %$k=2\pi\lambda$ is the characteristic spatial frequency of corrugations of %length scale $\lambda$ \cite{Jones2004JOPB}.
%One can therefore move further away from the wire, in order to reduce the %deleterious effects of the corrugations, albeit at the cost of reduced magnetic %gradient $\alpha$ and thus confinement. 
Routes to reduce the corrugations in the wave guides have been studied in  \cite{trebbia2007roughness,schumm2005matter}.
However, the mere fact that the shape of the atom chip potentials is defined by wire structures means that any imperfections in the wires cause defects and roughness in these potentials and make the single mode propagation over long distance very difficult to achieve.

%%%%%%%%%%%%%%%%%%%%%%%%%%%%%%%%%%%%%%%%%%%%%%%%%
%\subsection{Ultracold atoms-superconductors hybrid networks} 
%%%%%%%%%%%%%%%%%%%%%%%%%%%%%%%%%%%%%%%%%%%%%%%%%\label{hybrid} 

% \paragraph{Applications of atom chips}
% \Wolfn{Add Schmiedmayer / Josephson / transport / quantum simulation}
% Atom chip traps are very well suited for the generation and manipulation of Bose-Einstein condensates. \cite{Roux2008EEL}
% The first commercially available BEC machines \cite{ColdQuantaWWW} and many air- or space- bourn BEC experiments \cite{Rudolph2015NJOP} are indeed based on atom chips.
% One of the main draw-backs of atom-chip based waveguides is the inherent roughness of the potential \cite{Schumm2005TEPJD}.
% Increasing the distance of the atoms from the surface of the chip and reversing the electrical currents can reduce but not eliminate the resulting corrugation of the guiding potential \cite{Trebbia2007PRL}.
% This can then couple the forward momentum of an atom traveling along the waveguide to the transverse modes, thus causing heating and random phase shifts, thus severely limiting atomtronic devices such as guided matter wave interferometers.
%%%%%%%%%%%%%%%%%%%%%%%%%%%%%%%%%%%%%%%%%%%%%%%
\subsubsection{Adiabatic Magnetic Potentials}
%%%%%%%%%%%%%%%%%%%%%%%%%%%%%%%%%%%%%%%%%%%%%%%%
Adiabatic magnetic potentials offer a very interesting alternative to chip-based structures. They can be used to create a limited number of perfectly smooth trapping structures such as bubbles, rings and sheets.
They occur when  a radio-frequency field $(\vec{B}_\mathrm{rf})$ strongly couples magnetic hyperfine states and are readily described in  the dressed dressed-atom picture \cite{Tannoudji1977Dressed}.
When the radio frequency field is resonant with the magnetic field, i.e.\,$\omega_\mathrm{rf}=\omega_\mathrm{L}$, then the coupling strength can be expressed as the Rabi frequency
$
 \Omega_0 = g_\mathrm{F} \mu_\mathrm{B} B^\perp_\mathrm{rf} /\hbar\,,
$
where $B^\perp_\mathrm{rf}$ is the amplitude of the circularly polarized component of $\vec{B}_\mathrm{rf}$ that is orthogonal to $\vec{B}$ and couples the $m_\mathrm{F}$ states.
For an arbitrary detuning $(\delta=\omega_\mathrm{rf}-\omega_\mathrm{L})$ of the rf-field from the resonance the dressed potential can be expressed as
$	U(\vec{r})=m'_\mathrm{F}\hbar\sqrt{\delta^2(\vec{r})+\Omega_0^2(\vec{r})}$.
% $	U(\vec{r})=m'_\mathrm{F}\hbar\left[{\delta^2(\vec{r})+\Omega_0^2(\vec{r})\right]^{1/2}$.
% \begin{equation}
% 	U(\vec{r})=m'_\mathrm{F}\hbar\sqrt{\delta^2(\vec{r})+\Omega_0^2(\vec{r})}.
% \end{equation}
Note that the potential is equal to the non-dressed Zeeman states with $m'_\mathrm{F}=m_\mathrm{F}$ for $\delta \gg \Omega_0$, and inversely that it is equal to the non-dressed Zeeman states with $m'_\mathrm{F}=-m_\mathrm{F}$ for $\delta \ll -\Omega_0$.
% A sketch of the dressing can be seen in Fig.\,\ref{fig:Dress}.

% \begin{figure}
%  \includegraphics[width=0.7\columnwidth]{dressing}
%  \caption{Adiabatic potentials (solid lines) with their corresponding unperturbed Zeeman states (dashed). The two adiabatic potentials are split by twice the Rabi frequency: $U=2\hbar\Omega_0$. \label{fig:Dress}}
% \end{figure}

Let us examine the simple case of a magnetic quadrupole field, % (Eq.\,\ref{eq:B-Quadrupole}).
where the magnitude of the field increases linearly in all directions. In any direction starting from the center outwards, there is some point at which the rf field becomes resonant. 
The dressed field therefore forms an oblate bubble shaped trap, which has a radius of $ r_\rho=\hbar \omega_\mathrm{rf}/\left(\left|g_\mathrm{F}\right|\mu_\mathrm{B}\alpha\right)$ in the x-y plane and $ r_\mathrm{z}=\hbar \omega_\mathrm{rf}/\left(\left|g_\mathrm{F}\right|\mu_\mathrm{B}2\alpha\right)$ in the z-direction, where
$\alpha$ is the quadrupole field of gradient. 

The original idea was proposed by  \onlinecite{Zobay2001PRL} and first realized in \cite{Colombe2004EL}. A thorough review of these traps is found in \cite{Garraway2016JOPB,Perrin2017AIAMAOP}.

The dressed quadrupole field itself presents the problem that any homogeneous $B_\mathrm{rf}$ has one or two points on the bubble, where due to the projection of the rf onto the local quadrupole field the coupling field $B^\perp_\mathrm{rf}$ is zero, leading to Majorana spin-flip losses.
This can be avoided using a IP-type trap, where the magnetic field points predominantly in the direction of the $z$-axis.
In the absence of gravity, the quantum fluid can fill the entire bubble \cite{Sun2018PRA}. These hollow Bose-Einstein condensates are currently under investigation at the international space station \cite{Frye2021EQT}.
On earth however, gravity deforms the bubble-trap into something more akin to a cup, which can be exploited for 2D quantum gasses for its strong (weak) confinement in the vertical (horizontal) direction.
Using multiple rf-frequencies, multiple shells can be manipulated almost independently \cite{Bentine2017Species} and exploited for matter wave interferometry \cite{Mas2019NJOP}.

A dressed quadrupole trap can be used to create a matter wave ring simply due to the angular momentum of the trapped atoms and thus approaching a giant vortex state \cite{sherlock2011time-averaged,	guo2020supersonic}. %Note that FOOT demonstrated this already in 2011. Figure 7b
%%%%%%%%%%%%%%%%%%%%%%%%%%%%%%%%%%%%%%%%%%%%%%%%%%
%\subsubsection{Adiabatic Magnetic + Dipole}
%%%%%%%%%%%%%%%%%%%%%%%%%%%%%%%%%%%%%%%%%%%%%%%%%%
Alternatively, one can combine the rf-bubble trap with a red-detuned light-sheet, thus forming a ring-shaped trap \cite{morizot2006ring}. 

%%%%%%%%%%%%%%%%%%%%%%%%%%%%%%%%%%%%%%%%%%%%%%%%%
\subsubsection{Time-Averaged Adiabatic Potentials (TAAPs)}
%%%%%%%%%%%%%%%%%%%%%%%%%%%%%%%%%%%%%%%%%%%%%%%%%

\begin{figure}
\label{fig:TAAPCurrent}
 \includegraphics[width=1\columnwidth]{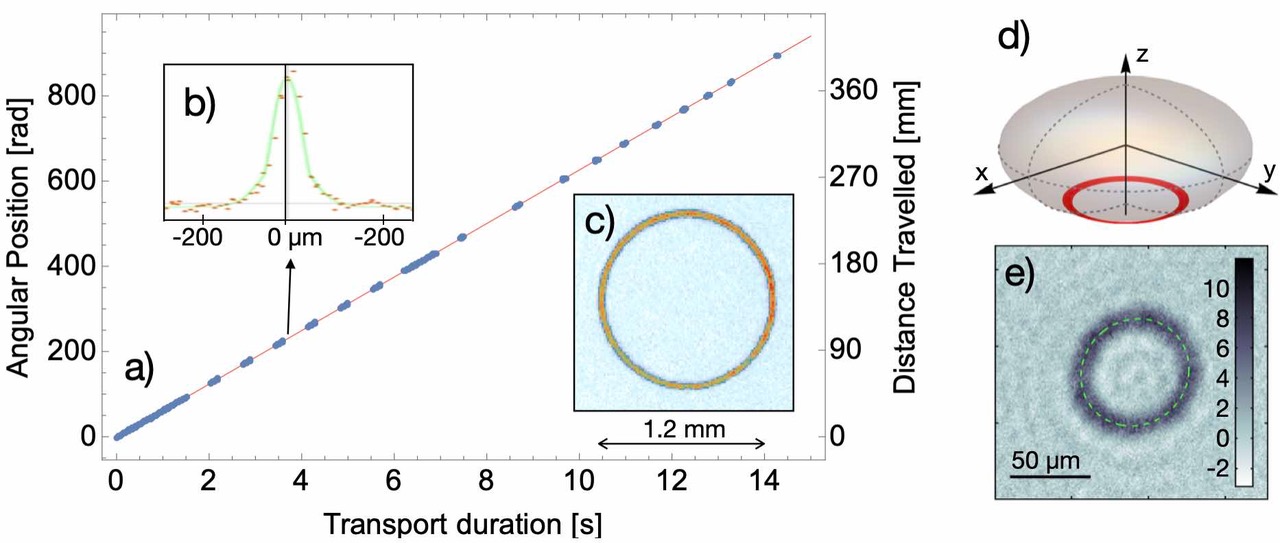}
 \caption{a) 
Long distance transport in a ring-shaped TAAP wave-guide. The plot shows the angular position of the condensate and thermal cloud during $14$\,s of transport in the matter-wave guide (blue dots) over a distance of more than $40$\,cm. The red line depicts the programmed trajectory of $2\pi 10$\,rad\,s$^{-1}$.
 b) The bi-modal distribution of
 the BEC after 4.1\,s of transport and a TOF expansion of 24\,ms, with the black
 arrow indicating to the relevant data point. See also \onlinecite{pandey2019hypersonic}. 
 c) A false-color absorption image of an annular condensate in a TAAP trap. d) A graphical illustration of atoms in a dynamically created ring
 trap and e) an absorption image of a BEC in it. Adapted from \onlinecite{guo2020supersonic}
 } 
\end{figure}

Highly versatile and controllable potentials in a great variety of perfectly smooth shapes can be created by combining the aforementioned adiabatic potentials with an oscillating, homogeneous magnetic field. 
If the Rabi frequency of the rf dressing field is large compared to the frequency of the oscillating field $\Omega_0 \gg \omega_\mathrm{m}$, the resulting trapping potential is the time-average of the adiabatic and modulation potentials \cite{lesanovsky2007time,navez2016matter}.
Starting with a simple quadrupole field and adding a vertically polarized rf-field, plus a vertical modulation field, one can generate a ring-shaped trap.
% , resulting in the trapping potential of the form 
% \begin{align}
% \label{eq:SimpleRingPotential}
% \begin{split}
%  V_{\text{r}}(r,z) =\, &\hbar \Omega_0 + \frac{1}{2}m{\omega_{\text{r}}^{2}}(r-R){^2} +\frac{1}{2}m{\omega_{\text{z}}^{2}}z^2 \\
% 	&-\frac{\delta }{2}m g R \cos\left({ \phi-\phi_0} \right),
% \end{split}	
% \end{align}
%  where $R$ is the radius of the ring, ${\omega_{\text{r}}}$ and ${\omega_{\text{z}}}$ are the radial and the axial trapping frequency, respectively. 
Typical values are 50\,Hz to 100\,Hz for the radial and axial trapping frequencies and 50\,$\mu$m to 1\,mm for the radius.
The ring can then be adiabatically turned into one or two coupled half-moon shaped traps simply by changing the polarization of the rf or modulation fields. 
Multiple concentric or stacked rings can be created by using more than one rf-frequency. 

The exact shapes of these traps and the barriers between them depend only on the amplitude and polarization of oscillating magnetic fields, which can be adjusted with extreme precision using standard electronics; making it possible to control the trapping potentials down to the picokelvin level.

Using a suitable choice of polarizations, it is also possible to trap two different spin states in identical effective potentials and even to manipulate them entirely independently \cite{navez2016matter}.
This technique might be exploited in an atom interferometer, where the atoms are placed in the TAAP in a single hyperfine state and exposed to a suitable microwave pulse. The hyperfine states of the resulting superposition can then be manipulated separately, making them sensitive to gravitation, for instance, and then recombining with a second microwave pulse - resulting in a highly sensitive interferometer \cite{navez2016matter}.
A similar scheme has been proposed for adiabatic ring-shaped potentials resulting from specially tailored magnetic fields 
 \cite{stevenson2015sagnac}.

Another feature of the TAAP rings is the extreme smoothness of these potentials \cite{pandey2019hypersonic}. 
Its shape is not determined by current carrying structures in proximity but by a quadrupole field and the frequency and amplitude of modulations fields, which are all generated by very distant coils. 
Therefore, any imperfections in the field-generating coils are exponentially suppressed on the size-scale of the trapping potentials. 
This is evidenced in Ref. \cite{navez2016matter}, where a Bose-Einstein condensate are transported at hypersonic speeds for a distance of 15\,cm without loss in spatial coherence.
%% MGB: I don't think the following sentence is a valid comparison. The Wang et al experiment looks at the loss of contrast in interference between wavepackets that accumulate phase (and phase gradients) as they propagate over different paths.
%% The observed loss of contrast is not blamed on guide roughness (it is most likely due to phase gradients arising from axial curvature in the guide. The TAAP experiment shows that the BEC remains a BEC i.e. there is little heating. But it doesn't measure the differential phase fluctutations important to interferometry and so, while 15cm propagation is impressive, comparisons to the interferometer coherence can't be made.
%% This coherence has to be compared to the atom-chip based interferometers, where the coherence is typically lost after a fraction of a millimeter \cite{wang2005atom}.
%It would be interesting to investigate the boundary between these two extreme regimes.
%

%%%%%%%%%%%%%%%%%%%%%%%%%%%%%%%%%%%%%%%%%%%%%%%%%
\subsection{Atom optical elements}
%%%%%%%%%%%%%%%%%%%%%%%%%%%%%%%%%%%%%%%%%%%%%%%%%
%%%%%%%%%%%%%%%%%%%%%%%%%%%%%%%%%%%%%%%
%\subsubsection{Intro}
%%%%%%%%%%%%%%%%%%%%%%%%%%%%%%%%%%%%%%%

% \Wolfn{I propose that after (or before) stating the different types of potentials that we can create, we describe what atom optical elements are needed in order to do atomtronics. The following come to mind: 3D Traps, trap/beam splitters, waveguides, multiply connected traps (ring).\\ Alternatively, I will include a description of the requirements for a waveguide in the chip and TAAP sections, beam splitter in the TAAP.}

In this section, we outline the types of potentials and the optical elements that have been designed to guide the matter-wave in atomtronic circuits.

\subsubsection{Waveguides}

The fabrication of one dimensional guides is important in the atomtronics context, both to control the circuit functionalities and to explore quantum effects in fundamental physics. 
The coherent regime needs to consider both the tightness of the confinement and any displacement or roughness of the waveguide transverse to the direction of propagation. Operating in the strict one-dimensional regime requires the transverse frequencies to be much larger that both the chemical potential and the temperature of the gas, as well as the kinetic energy originating from the current flowing in the system. This can be achieved using tight optical confinement from optical lattices \cite{bloch2008many} or projected wires \cite{krinner2017two-terminal}. %In case the strict one-dimensional regime is not reached, ballistic flow is nevertheless encountered provided the current flows adiabatically \cite{glazman1988reflectionless} across regions with varying transverse confinement.
Optical lattices with a typical lattice spacing of few microns have been realized \cite{rubinsztein-dunlop2016roadmap}. 
In such structures, in which the cold atoms can tunnel between the lowest Bloch bands of the adjacent wells, the low temperatures matter-wave effective dynamics is one dimensional.

%\Luigin{Add conditions for one dimensional guides/rings \Wolf{Jean-Philippe: mention energy scales.}}

% \JeanPhilippen{Here is the place where one should mention the non-linear beam-splitters such as the atom-chip experiment of Treutlein and the Oberthaler squeezing ? \Wolf{I agree... and will implement it}}

A rapid modulation of the strength of the transverse confinement or a bend in the waveguide couples the forward motion of the atoms to oscillations of the condensate via its chemical potential. 
It also causes a shift of the potential energy of the bottom of the waveguide, which can induce scattering, although it is possible to shape a \added[id=WK]{slow} bend to avoid this effect \cite{delcampo2014bent}. 
As pointed out in section \ref{sec:AtomChips}, this regime is very difficult to achieve with atom chips.
Excitationless matter-wave guides have been demonstrated using single-beam optical dipole beams over distances of 3.5\,mm \cite{mcdonald2013optically} and using TAAPs over distances of 40\,cm for thermal clouds and 15\,cm for BECs \cite{pandey2019hypersonic}. Waveguide bends and junctions created with painted potentials have demonstrated excitation probabilities of less than 8\% \cite{ryu2015integrated}.

The requirements for precision interferometers are rather stringent. 
Care must be taken to ensure that the superposition state traverses the interferometer adiabatically and that the trap-induced energy difference between the two paths is extremely well controlled \cite{Zabow2004PRL,Kreutzmann2004PRL}.
For propagating matter wave interferometers this results in extreme requirements on corrugations of the waveguides, since any small lateral deviation tends to couple the forward motion to transverse modes -- thus destroying the coherence of the superposition state.
In the absence of such coupling the interferometer are expected to be able to operate using multiple transverse modes concurrently \cite{andersson2002multimode}.

\subsubsection{Ring traps \label{sec:rings}}

% \Luigin{I suggest we make a section on ring traps,\Wolf{OK, working on it}}
 % \cite{hopkins2004proposed} MAGNETO ELECTROSTATIC 

% first ring sauer2001storage

Ring traps are the simplest spatially closed  atomtronic circuits. 
They normally consist of a tight harmonic confinement in the vertical and horizontal directions and no confinement along the azimuthal direction. 
The potential for a ring of radius $\rho_0$ can be written near the trap minimum as $U = \frac{1}{2} m \omega_\rho^2 (\rho-\rho_0)^2 +\frac{1}{2} m \omega_z^2 z^2$, with $m$ being the mass of the atoms and $\omega_\rho$ and $\omega_z$ are the harmonic trapping frequencies in the radial and vertical directions respectively.
There are two distinct regimes of interest for ring traps: One where the radius of the ring is small enough for the energy or time-scale of the excitations along the circumference of the ring to become important, and one, where the ring is to be viewed more like a circular waveguide. 

Early demonstrations focused on atom propagation in large ring-shaped waveguides. 
Large-scale magnetic traps have been demonstrated, where the shape of the ring is directly defined by the field generating wires, either using microfabricated `atom-chips' \cite{sauer2001storage} or large wire structures \cite{gupta2005bose-einstein, arnold2006large} and inductively coupled rings \cite{pritchard2012Demonstration}.
For polar molecules electrostatic ring-traps have also been demonstrated \cite{crompvoets2001prototype}. 

Small-diameter ring traps can \deleted[id=WK]{also} be generated using purely optical dipole potentials. The first toroidal BEC is created with a painted potential \cite{henderson2009experimental}. Static approaches to creating rings with optical potentials include using a light-sheet in combination with a Laguerre-Gauss beam \cite{wright2000toroidal,ramanathan2011superflow,moulder2012quantized}.
Persistent currents in spinor condensates are detected in \cite{beattie2013persistent}.
In this case, the imperfection in the potential often do not lead to adverse effects since the superfluid nature of the flow smooths them over.
Examples include studies of super-currents \cite{moulder2012quantized, eckel2014hysteresis, ryu2014creation} and Josephson junctions \cite{wright2013Driving, ryu2013experimental}.
 
% MGB: The smoothness discussion in the following paragraph has already been presented above.
% For coherent transport of Bose-Einstein condensates and matterwave interferometry, however, imperfections are of major concern.
% Atomtronic circuits manufactured by the aforementioned methods can produce almost arbitrary structures, which means in turn that corrugations of the waveguide can exists and therefore to some extent always will exist.
% In contrast to this, Time-Averaged Adiabatic Potentials \cite{pandey2019hypersonic,sherlock2011time-averaged,lesanovsky2007time} and dynamic ring traps \cite{guo2020supersonic} are defined by dressing atoms in a simple magnetic %quadrupole field (see Fig.\,\ref{fig:TAAPCurrent}). 
% It is practically impossible for these methods to produce a modulation in the potential, which exceeds an angular frequency of $4 pi$.
% The extreme smoothness of these waveguides has allowed the hypersonic propagation of BECs over distances extending 15\,cm \cite{pandey2019hypersonic}, and thus show great promise for guided matter wave interferometry.
% Ring-shaped atomtronic circuits have already resulted in many interesting experiments. 

Optical ring lattices suitable for trapping cold and quantum degenerate atomic samples have been generated with Laguerre-Gauss laser beams incident on SLM or manipulated by an Acousto-Optical Modulator \cite{franke-arnold2007optical,amico2014superfluid, henderson2009experimental}. 
By applying computer assisted optimization algorithms using absorption images of BECs in the dipole potentials, smooth potentials of rings radii and lattice spacings in the range of $100\mu m$ and $30 \mu m$ respectively have been demonstrated. 
DMD generated rings lattices of $\sim 40\mu m$ and lattice spacing of $\sim 4 \mu m$ have been achieved \cite{gauthier2016direct}. 
However, in order to achieve any appreciable tunneling among the lattice wells, such numbers need to be scaled down further.

Finally, ultra-smooth ring-shaped waveguides  based on Time-averaged Adiabatic Potentials (TAAP)s \cite{lesanovsky2007time} have recently been demonstrated. They support coherent, lossless transport of matterwaves over very macroscopic distances (14\,cm) even at hypersonic speeds \cite{pandey2019hypersonic}.

%%%%%%%%%%%%%%%%%%%%%%%%%%%%%%%%%%%%%%%%%%%%%%%%%%%%%%%%%%%%%%%%%%%%%%%%%%%%%%
\subsubsection{Barriers and  beam splitters}
\label{sec:BarriersAndbeam splitters}
%%%%%%%%%%%%%%%%%%%%%%%%%%%%%%%%%%%%%%%%%%%%%%%%%%%%%%%%%%%%%%%%%%%%%%%%%%%%%%

The terms ``barrier" and ``beam splitters" are primarily distinguished by their intended use rather than by underlying function or principles. Borrowing the term from optics, beam splitters are familiar elements, particularly in the context of interferometers, used for coupling a pair of system modes. Barriers are commonly used to define spatially distinct regions, for example, having different potential structures, temperatures, and chemical potentials as is the case in triple-well transistors, for example \cite{caliga2016principles}. Beam splitters have been implemented using both time-dependent and time-independent potentials, whereas barriers are typically implemented with time independent potentials. 

Early work on atomtronic beam splitters split magnetically guided thermal atoms utilizing ``Y"- or ``X"-shaped conductor junctions, resulting in multi-mode splitting \cite{Cassettari.2000, Muller.20000rh}. Coherent beam splitting on an atom chip has been carried out using Bragg diffraction from an optical lattice that is exposed to the atoms in a double-pulsed manner \cite{Diot.2004, wang2005atom}. Coherent splitting of stationary Bose condensates produced on atom chips has also been carried out utilizing radio-frequency fields that provide and elegant means of evolving in time a single magnetically generated potential well into a double well and vice-versa \cite{schumm2005matter,Hofferberth.2006,Kim.2017}. Of note, the earlier successes in coherent beam splitting utilized time-dependent potentials. 

In the framework of atomtronic circuitry there is particular interest in beam splitters that are spatially fixed and time independent. The use of ``painted potentials" \cite{henderson2009experimental} enabled coherent splitting of a propagating condensate in ``Y" junction optical waveguides \cite{ryu2015integrated} while beam splitting in crossing optical waveguides has also been carried out using an optical Bragg grating produced by a pair of interfering laser beams located at the waveguide junction \cite{Guarrera.2017}.

In contrast to the coherent splitting, that is achieved with waveguides coupled by spatial proximity or through optical gratings, barriers act more like the mirrors and beam splitters of optical systems. Barriers produced using projected blue-detuned laser beams feature a smooth Gaussian profile. Coherent splitting occurs due to tunneling and quantum reflection for atoms with energies below or above the top of the barrier, respectively, within an energy range proportional to the inverse curvature of the barrier at its top \cite{cuevas2017molecular}. 
Matter wave propagation across arbitrary arrangements of barriers can be numerically calculated utilizing the impedance method \cite{khondker1988a, Gutierrez-Medina.2013}, a technique that borrows wave propagation techniques from electromagnetic transmission line analysis.% For example, two Gaussian barriers separated by a specified distance form the matter wave equivalent of a Fabry-Perot cavity, which can be used to select a narrow spectrum of particle energies \cite{Dinardo2018}. Indeed, one can design the matter wave equivalent of multilayer dielectric coatings, partial reflectors, 50/50 beam splitters and so on.

Projected optical atomic potentials that affect the center-of-mass 
motion through AC Stark shifts are limited in size-scale by 
the wavelength of the projected light. 
Barriers having sub-wavelength size scales down to less than 
$50 nm$ have been demonstrated using the nonlinear response of 
the dark state of a three-level system \cite{lkacki2016nanoscale,Wang2018}. 

Regarding the large variety of approaches in barrier and beamsplitter implementation, it is not evident that a single approach has emerged as best suited for atomtronic systems. Rather, the optimum approach is purpose-dependent. What has emerged, however, is that it has proved quite difficult to achieve coherence-preserving barriers or beam splitters using purely magnetic approaches. Either all optical or hybrid magnetic and optical or radio-frequency systems have met with good success.

%%%%%%%%%%%%%%%%%%%%%%%%%%%%%%%%%%%%%%%%%%%%%%%%%%%%%%%%%%%%%%%%%%%%%%%%%%%%%%
% \subsubsection{Lenses}
% \label{sec:Lenses}
%%%%%%%%%%%%%%%%%%%%%%%%%%%%%%%%%%%%%%%%%%%%%%%%%%%%%%%%%%%%%%%%%%%%%%%%%%%%%%

%Arguably the most fundamental element of optics is the lens. 
%As opposed to their photonic counter part, lenses in atomtronic circuits act on longitudinal degree of freedom. 
%They can be used, e.g., in order to control the dispersion of a matterwave or to delta-kick cool the atoms. 
%A gravito-magnetic matterwave lens has recently been demonstrated by %\onlinecite{pandey2021atomtronic}.

%%%%%%%%%%%%%%%%%%%%%%%%%%%%%%%%%%%%%%%
%\subsubsection{matter wave guides}
%%%%%%%%%%%%%%%%%%%%%%%%%%%%%%%%%%%%%%%%
%
%Magnetically Guided Atomic Beam \cite{Lahaye2004PRL}
%Magnetic atom optics: mirrors, guides, traps, and chips for atoms "ideas may be brought together to achieve integrated atom optics on a chip" \cite{Hinds1999JOPD,Folman2000PRL}	
%
%- BEC \cite{Sinclair2005PRA} %Bouchoule on reversing current to suppress roughness 
%
%- Interference \cite{Shin2005PRA}
%
%- Double well \cite{Sinclair2005PRA,Shin2005PRA,Schumm2006QIP,Lesanovsky2006PRA}
%
%
%- Inductive \cite{Vangeleyn2014JOPB}
%
%
%- Rings
%
%- Beam Splitter for Guided Atoms \cite{Cassettari2000PRL}.
%
%- Obstacles / Barriers \cite{navez2016matter}
 %

%

%\input{measures}

 \section{Coherent effects in mesoscopic matter-wave circuits}
 \label{CoherentMesoscopics}

Like their electronic counterparts \cite{cuevas2017molecular}, atomtronic devices can operate in a regime where quantum interferences play a dominant role. Such a coherent regime is achieved in situations where the typical transport scale, such as the circumference of a ring trap or the length of a mesoscopic section, is larger than the typical decay length of the particles' correlation function. Phenomena such as Aharonov-Bohm interferences, Bragg reflection on periodic structures or Anderson localization emerge in the transport properties. Quantum coherent transport is deduced from properties of the Hamiltonian describing the atoms inside the conductor, identical to coherent wave propagation in complex media encountered in electromagnetism.

Operating in this regime typically requires low enough temperatures: for Fermi gases, the relevant length scale is $\frac{\hbar v_F}{k_B T}$, where $v_F$ is the Fermi velocity, and for thermal gases it is the thermal de Broglie wavelength. For bosons, the emergence of the condensate below the critical temperature ensures coherence at arbitrarily large distances, making them particularly well suited for the study of coherent transport. The long wavelength dynamics is then efficiently described by superfluid hydrodynamics (see Sec.\ref{subsec:bosons}). Many-body fermionic particles are discussed in Sec{subsec:fermions}. Coherence properties are also affected by many-body effects such as the decrease of quasi-particle lifetime and quantum fluctuations. Furthermore, coherence can be reduced by noise, spontaneous emission from lasers or other external disturbances.

%Many-body effects also affect the coherence properties, 
%for example by decreasing of the quasi-particle lifetime, or more dramatically upon approaching a Mott insulating state. 

In this section, we bridge microscopic models to simple matter-wave circuits. The various model Hamiltonians 
%Hamiltonian models provide 
describing coherent quantum fluids with different features are introduced. We then 
%will 
focus on the persistent current in ring-shape circuits providing both an important figure of merit for the system's coherence and an elementary building block for atomtronic circuits. Finally we present the two-terminal quantum transport and illustrate the specific features emerging from the coherent quantum dynamics.
%The coherent dynamics emerges in specific features of the two-leads quantum transport.
\subsection{Model Hamiltonians}
\label{hamiltonians}

% \JeanPhilippen{I made few changes to the formulation}
%\JeanPhilippen{I propose to also include the tunneling Hamiltonian %for point contacts.}
%Here we introduce the models describing the %quantum dynamics in the matter-wave %circuits. 
The many-body Hamiltonian describing $N$ interacting quantum particles of mass $m$, subjected to an effective magnetic field described by the vector potential ${\mathbf A}$ and confined in the potential $V_{ext}$ reads
\begin{eqnarray}\begin{split}
\label{manybodyhamiltonian}
H= \int d {\mathbf r} \Psi^\dagger({\mathbf r}) \left [ {{1}\over{2m}} \left (-i \hbar \nabla+ {\mathbf A} ({\mathbf r}) \right )^2 +V_{ext}(\mathbf{r}) \right ] \Psi({\mathbf r}) 
% \nonumber 
 \\ + {{1}\over{2}} \int d {\mathbf r} d {\mathbf r'} \Psi^\dagger({\mathbf r}) \Psi^\dagger({\mathbf r'}) v({\mathbf r} - {\mathbf r'} ) \Psi({\mathbf r}) \Psi({\mathbf r'}) 
\end{split}\end{eqnarray} 
where $\Psi^\dagger({\mathbf r}) $ and $\Psi({\mathbf r})$ are field operators creating or annihilating a bosonic or fermionic particle at the spatial position ${\mathbf r}$ \cite{mahan2013many} and $v({\mathbf r} - {\mathbf r'} )$ is the two-body inter-particle interactions. We assume contact interactions $v({\mathbf r} - {\mathbf r'} )= g \delta({\mathbf r} - {\mathbf r'} ) $, with $g=4\pi \hbar^2 a_s/m$ and $a_s$ the $s$-wave scattering length. 
${\mathbf A}$ is an effective gauge potential, which plays a crucial role in the description of currents in spatially closed geometries.
%If not explicitely stated otherwise, in this colloquium closed spatial architecture will be considered. 
Both lattice and continuous systems are relevant for atomtronic circuits. The quantum many-body theories will be presented mostly for the one-dimensional case. They will be used to describe quantitatively tightly confined geometries, such as quantum wires and point contacts, but also capture qualitatively the physics of extended systems along the transport direction. Given their relevance for specific atomtronic circuits, the Gross-Pitaevskii mean field theories will also be discussed for higher dimensions.

\subsubsection{Bosons}
\label{subsec:bosons}
In this case the field operators obey the commutation relations $\left [\Psi({\mathbf r}), \Psi^\dagger({\mathbf r'})\right ]=\delta ({\mathbf r}-{\mathbf r'}) $. For a recent review on one-dimensional bosons see \cite{RevModPhys.83.1405}. We start with a lattice 
theory describing atoms localized in potential wells centered in $N_s$ sites, and expand the field operators in Wannier functions, assumed to be a good basis of eigenfunctions of separated local potential wells: $\Psi({\mathbf r})=\sum_{j=1}^{N_s} w({\mathbf r}-{\mathbf r_j}) a_j$, in which the operators $\hat a_j$ create a single bosonic particle at the site $j$, $ [a_i,a^\dagger_j ] =\delta_{ij}$. With the above expression of $\Psi({\mathbf r})$, the many body Hamiltonian can be recast to the Bose-Hubbard Model (BHM)
\begin{equation}\label{BHM}
\mathcal{H}_\text{BH}=\sum_{\langle i,j \rangle}^{N_s}\left[ -J \, \left( {a}_{j}{a}_{j+1} + {a}_{j+1}{a}_{j}\right)+ \frac{U}{2}\nn{}{j}\left(\nn{}{j}-1\right)\right],
\end{equation}
in which we assumed that only atoms in nearest neighbor local wells can appreciably overlap. % and that $V(({\mathbf r}) - ({\mathbf r'}) =\delta({\mathbf r} - {\mathbf r'}) $ 
A ring geometry is assumed, such that $\cn{a}{N_s+1}=\cn{a}{1}$. The parameters in the Hamiltonian are the hopping amplitude $J=\int d\textbf{r} w^{*}(\textbf{r}-\textbf{r}_i) \left [ {{1}\over{2m}} \left (-i \hbar \nabla+ {\mathbf A} ({\mathbf r}) \right )^2 +V_{ext} \right ] w(\textbf{r}-\textbf{r}_{i+1})$ and interaction strength $U=\pi a_s \int d \textbf{r} |w(\textbf{r})|^4/m$.
%, $a_s$ being the atom-atom scattering length.
The Hamiltonian (\ref{BHM}), originally introduced as a lattice regularization of the continuous theory of bosonic fields \cite{haldane1980solidification}, provides a paradigmatic model to study Mott insulator-superfluid quantum phase transitions \cite{fisher1989boson}. The BHM is extensively used in mesoscopic physics \cite{fazio2001quantum}. The conditions for the realization %relevance
of the 
the BHM in %for 
cold atoms systems are identified 
%is recognized 
in \cite{jaksch1998cold}, and since then it provides an important scheme in the cold-atoms quantum technology \cite{bloch2008many}. For neutral matter, the vector potential $\textbf{A}(\textbf{x},t)$ provides an artificial gauge field \cite{dalibard2011colloquium,goldman2014light-induced} - see Sect.\ref{persistent}.
For sufficiently smooth $\textbf{A}(\textbf{x},t)$ on the atomic scale, the 
%synthetic
gauge field can be absorbed in the Wannier functions 
$\tilde{w}(\textbf{r}-\textbf{r}_i)=e^{-i \Lambda(\textbf{r},t)} w(\textbf{r}-\textbf{r}_i)\approx e^{i \Lambda(\textbf{r}_i,t)} w(\textbf{r}-\textbf{r}_i)$ with $\Lambda(\textbf{r},t)=\int_{\textbf{r}_0}^{\textbf{r}} \textbf{A}(\textbf{r},t)d\textbf{r}$,where $\textbf{r}_0$ is an arbitrary lattice site. 
Therefore the hopping parameter results 
$ %\begin{equation}
\displaystyle{J=e^{i\Phi}J_0 \;,  \Phi =\int_{\textbf{r}_i}^{\textbf{r}_{i+1}} \textbf{A}(\textbf{r},t)d\textbf{r}
}
$. %\end{equation}
The procedure of absorbing the effects of the gauge field into the hopping matrix element is called Peierls substitution \cite{peierls1933theorie,essler2005one}.

In the limit of a large average number of particles per site ${\nu=N/N_s\gg1}$
%\Annan{$N_\text{p}$ replaced by N},
%small on-site number fluctuations $\Delta\nn{}{j} \ll N_s$ and 
%small on-site particle number fluctuations around ${\overline{Q}}$ and 
%interaction strengths in the range ${4J/N_s \ll U \ll 4JN_s}$ \cite{huber2008amplitude}, 
the Bose-Hubbard Hamiltonian effectively reduces to the Quantum Phase Model (QPM) \cite{fazio2001quantum}
%: $a_j\sim e^{i\hat{\phi}_j}$. 
\begin{equation} 
%\mathcal{H}_\text{QP}= - \sum_{<i,j>}^L 2 \tilde{J} \cos (\hat{\phi}_i-\hat{\phi}_{j} - \Phi_{i,j}) + \sum_{j=1}^L P_j(t) \hat{Q}_j + \frac{U}{2} \sum_{j=1}^L \hat{Q}^2_j+\text{const}\, ,
\mathcal{H}_\text{QP}= -2 J_\text{E}\sum_{\langle i,j \rangle}^{N_s} \Big[ \, \cos (\hat{\phi}_i-\hat{\phi}_{j} -\Phi ) + \frac{U}{2} \sum_{j=1}^{N_s}\hat{Q}^2_j \Big], 
\label{Eq:QP}
\end{equation}
where 
%$\tilde{J}=\overline{Q}J$ is the rescaled nearest-neighbor coupling and 
$J_\text{E}=J N_\text{s}$, $\hat{Q}_j = \nn{}{j}-N/N_\text{s}$ is the on-site particle number fluctuations and $\phi_j$ the (Hermitian) phase operators~ \cite{amico2000time,amico2000algebraic}. The operators satisfy the commutation relations ${[\hat{\phi}_i,\hat{Q}_j] = \mathrm{i}\hbar \delta_{ij}}$.

In the limit of small filling fractions $\nu=N/N_s\ll 1$
%\Annan{already defined as $N_s$}, 
the lattice Hamiltonian Eq.(\ref{BHM}) leads to the Bose-gas continuous theory. This 
statement holds true since the filling is proportional to the lattice spacing $\Delta$: $\nu=D \Delta$, with $D$ being the particle density.
In order to have a well defined result in the continuous limit $\Delta \rightarrow 0$, the bosonic operators must be rescaled: $\hat a_i=\sqrt \Delta \Psi (\mathbf{r}_i)$,
 $\hat n_i= \Delta \Psi^\dagger (\mathbf{r}_i)\Psi (\mathbf{r}_i) $, $\mathbf{r}_i= \mathbf{i} \Delta $.
The BHM reads as \cite{korepin1997quantum}:
$H_{BH}=t\Delta^2 {\cal{H}}_{BG}$,
$%\begin{equation}
{\cal H}_{BG}=\int d\mathbf{r} \left [(\partial_\mathbf{r} \Psi^\dagger)(\partial_\mathbf{r} \Psi)
+ c \Psi^\dagger\Psi^\dagger\Psi\Psi \right ]
%\label{bose-gas}
$, %\end{equation}
with $c=U/(t\Delta)$ \cite{amico2004universality}. This coincides with Eq.~(\ref{manybodyhamiltonian}) where $c=m g/\hbar^2$.
We note that while the procedure is valid for any $U$, the 
attractive case demands smaller values of $\Delta$ for the actual mapping of the spectrum \cite{oelkers2007ground-state}. This feature is due to formation of quantum analog of bright solitons \cite{naldesi2019rise}.
%\note{WK: Luigi, Please check citation.  ArXiv number has other title online}. 

%In the mean-field regime, where one sets $\langle \Psi(\mathbf{r})\rangle=\Phi(\mathbf{r})$, holding at weak interactions, we recover the Gross-Pitaevskii functional, $E_{GP}[\Phi,\Phi^*]=\int d\mathbf{r} \frac{\hbar\2}{2m}\left [(\partial_\mathbf{r} \Phi^*)(\partial_\mathbf{r} \Phi) + V_{ext}(\mathbf{r})|\Phi(\mathbf{r})|^2
% + g |\Phi(\mathbf{r})|^4 \right ]$ where we have restored dimensions and %included the external trap potential. This leads to the Gross-Pitaevskii equation upon functional differentiation.

%The Hamiltonian in first quantization can be obtained by solving the spectral equation for ${ H}_{BG}$: ${H}_{BG} |\psi(\mathbf{\lambda}) \rangle = {\cal E} |\psi(\mathbf{\lambda}) \rangle$, with $|\psi({\mathbf{\lambda}}) \rangle=\int d {\mathbf r} \chi({\mathbf r}|{\uathbf{ \lambda}}) \Psi^\dagger (\mathbf{r}_1)\dots \Psi^\dagger (\mathbf{r}_N)|0\rangle$, being $\mathbf{\lambda}\doteq \{\mathbf{\lambda}_1\dots\mathbf{\lambda}_{N_p}\}$ and $\mathbf{r}\doteq\{ \mathbf{r}_1 \dots \mathbf{r}_{N_p}\}$. Finally, it can be proved that $\chi({\mathbf r}|\mathbf{\lambda})$ is indeed eigenfunctions of 
% ANNA: too technical for the colloquium. Ref is enough

The many-body Hamiltonian arising from ${\cal{H}}_{BG}$in first quantization is known as the Lieb-Liniger model and reads 
\begin{equation}
{\cal H}_{LL}=\sum_{j=1}^{N_p} {\frac{\hbar^2}{2m}} \left ( -i \frac{\partial}{\partial x_j} - {{\Phi}\over{2 \pi N_s}} \right )^2 +g\sum_{ 1\le j <k <N_p} \delta (x_j-x_k) \,.
\label{delta}
\end{equation}
We note that, despite the Bose-Hubbard model is not integrable \cite{choy1982failure,amico2004universality,dutta2015non}, the 1D Bose gas is, with exact solution given 
%In this case, the first quantized Hamiltonian 
%Eq.(\ref{delta}) is known as the 
by Lieb and Liniger using the Bethe Ansatz \cite{LiebLiniger}.

%The Hamiltonian \eqref{BHH} commutes with the total number of particles $N$:
%
%\begin{eqnarray}
%\label{BHH}
%\hat{N}=\sum_{i=0}^{L} \hat{n}_i \ \ \ \ \ \ \ \left[ \hat{H}(U), \hat{N} \right] = 0,
%\end{eqnarray}
%
%therefore it can be diagonalized separately in every sector that has a well defined number of particles.
With a fully factorized (not-entangled) Ansatz for the many-body wavefunction
%$\chi_{GS}({\mathbf r}|{\mathbf{ \lambda}})=(1/N)\prod_j^N \phi(\mathbf{r}_j)$, 
$\Psi_{GS}({\mathbf{ r}_1,...\mathbf{r}_N})=(1/N)\prod_j^N \phi(\mathbf{r}_j)$
the dynamics entailed by the %Bose gas THAT ONE HAS A=0 
Hamiltonian (\ref{manybodyhamiltonian}) reduces to the Gross-Pitaevskii equation \cite{dalfovo1999theory,leggett2006quantum,calogero1975comparison} 
%\Wolfn{There is a problem in the formula %too many "]."}
\begin{eqnarray}
i\hbar \partial_t\phi(\mathbf{r},t)& = &\left[\frac{\hbar^2}{2m} \left(- i \nabla - \mathbf{A}\right)^2 + V_{ext}(\mathbf{r}) \right. \nonumber \\
& & \mbox{\hspace{1cm}} +g N\left |\phi(\mathbf{r})|^2 \right ] \phi(\mathbf{r}) \,,
\label{GPE}
\end{eqnarray}
in which we restored the 3D character of the system since many relevant applications of the GPE occur in circuits of higher dimensionality (e.g. toroidal confinements). 
%Connection between GPE solution and Lieb-liniger .
We note that $\sqrt{N} \phi(\mathbf{r}) $ coincides with $\langle \Psi({\mathbf r},t) \rangle$, defined by the mean field approximation %description which assumes REDUNDANT
%$\|\Psi({\mathbf r},t) - \langle \Psi({\mathbf r},t) \rangle \| \ll 1$ Psi is dimensionful so this does not work
%$\Psi({\mathbf r},t) \simeq \langle \Psi({\mathbf r},t)$
 of the Heisenberg equations of the motion
stemming from the Hamiltonian 
Eq.~(\ref{manybodyhamiltonian}).
The reduction of the quantum many body problem to the GPE dynamics is well justified in the dilute regime, i.e. when $a_s^3 \rho \ll 1$ (see \cite{lee1957eigenvalues} for corrections).
 In 1D, taking $A=0$ and $V_{ext}=0$ the Gross-Pitaevskii equation is integrable, with solitonic solutions \cite{faddeev2007hamiltonian}. Eq. (\ref{GPE}), recast in amplitude phase representation $\phi=\sqrt{n} e^{i \theta}$ gives rise to the superfluid hydrodynamics equations for the condensate density $n$ and the phase $\theta$ \cite{dalfovo1999theory}.

%\begin{equation}
% \hbar \partial \langle \Psi({\mathbf r},t) \rangle= \left [ \left (-i \hbar \nabla+ {\mathbf A} ({\mathbf r}) \right )^2+V_{ext}+g |\langle \Psi({\mathbf r},t) \rangle|^2 \right ]\langle \Psi({\mathbf r},t) \rangle
%\end{equation}

\subsubsection{Fermions}
%\Annan{I suggest to say SU($\kappa$) here because $N$ is the %particle number everywhere}
Here, we refer to a gas of fermions with 
%$N$ 
$\kappa$ components or colours.  In this case, the field operators are characterized by the spin label $\alpha=\{ 1,\dots \kappa\}$. They obey the anticommutation rules: $\left \{\Psi_\alpha({\mathbf r}), \Psi_{\alpha'}^\dagger({\mathbf r'})\right \}=\delta_{\alpha,\alpha'}\delta ({\mathbf r}-{\mathbf r'}) $.
By employing a similar derivation as %logic
described above for the bosonic case, one can obtain the generalization of the Hubbard model.  If the physical parameters of the system, like interaction or trapping potentials, turns out independent by the colour then $\kappa$-components fermions are known as SU($\kappa)$ fermions. The Hamiltonian for SU($\kappa$) fermions in a ring lattice pierced by an effective gauge field reads \cite{capponi2016phases}
\begin{align}\begin{split}
\label{eq:1}
\mathcal{H}_{\textrm{SU}(
\kappa)} = -J\sum_{j=1}^{N_s}\sum\limits_{\alpha = 1}^{\kappa}\big (e^{\imath \Phi}c^{\dagger}_{\alpha ,j}c_{\alpha ,j+1} + \textrm{h.c.}\big ) 
\\
+ 
%\frac{U}{2}\sum\limits_{j}n_{j}(n_{j}-1) \nonumber
U\sum\limits_{\alpha\neq \alpha'j} {n_{\alpha,j}n_{\alpha',j}} 
\end{split}\end{align}
where $c_{\alpha,j}^{\dagger}$ creates a fermion at the site $j$ of a $d$-dimensional lattice with spin component $\alpha$, %$n_{j} = \sum_\alpha c_{\alpha, j}^{\dagger}c_{\alpha,j}$
$n_{\alpha, j}= c_{\alpha, j}^{\dagger}c_{\alpha,j}$
is the local number operator for site $j$ and spin component $\alpha$. 
The parameters $J$ and $U$ account for the hopping strength and on-site interaction respectively. They can be expressed in terms of integrals of Wannier functions as discussed for the bosonic case. 
%(The size of the system is assumed to be large enough such that the Peierls substitution is well defined \cite{Peierls}. The interaction of the magnetic field with %the electron spin is neglected.)
For $\kappa=2$, \label{eq:1}
provides a paradigmatic framework to address the physics of itinerant  electrons,
in $d$-dimensional lattice \cite{hubbard1963electron,gutzwiller1963effect,kanamori1963electron}. See \cite{baeriswyl2013hubbard,mielke2015hubbard,mahan2013many} for more recent references.
Systems of two spin components \cite{jordens2008mott}, and more recently, of $\kappa$ components fermions \cite{pagano2015strongly,cappellini2014direct} have been experimentally realized with the cold atoms quantum technology. 
For $\kappa=2$, the Hamiltonian (\ref{eq:1}) is integrable by Bethe Ansatz for any values of system parameters and filling fractions $\nu=N/L$ \cite{lieb1968absence}. For $\kappa>2$, the Bethe Ansatz integrability is preserved in the continuous limit of vanishing lattice spacing, (\ref{eq:1}) turning into the Gaudin-Yang-Sutherland model describing SU($\kappa$) fermions with delta interaction \cite{sutherland1968further}; such regime is achieved by (\ref{eq:1}) in the dilute limit of small fillings fractions. Bethe Ansatz solutions allows the precise understanding both of the ground state and the nature of excitations of the system. The corresponding Hamiltonian reads:
\begin{align}\begin{split}
\label{GYS}
H_{GYS}= &\sum_{\alpha=1}^\kappa \sum_{j=1}^{N} {\frac{\hbar^2}{2m}} \left ( -i {{\partial}_{x_{j,\alpha}}-{{\Phi}\over{2 \pi N_s}}}\right )^2 \\ 
&+ g \sum_{1\le i<j\le N} \sum_{\alpha,\beta=1}^\kappa \delta (x_{i,\beta}-x_{j,\alpha})
\end{split}\end{align}
% if we put hbar and m in the kinetic part, the interaction constant is g. c has the dimensions of a wavevector
Another integrable regime of (\ref{eq:1}) is obtained for $\langle \sum_\alpha n_{\alpha,j}\rangle =1\forall j$ and large repulsive values of $U\gg t$ for which the system is governed by the $SU(\kappa) $ antiferromagnetic Sutherland model \cite{sutherland1975model,capponi2016phases,guan2013fermi}. 
 %{\it i)} in the the continuous limit and {\it ii)} for a filling of $n=1/N_{p}$ ($N_{p}=L$) i.e. one atom per site in the presence of large repulsive values of $U$
In the intermediate interactions and intermediate fillings, the model $(\ref{eq:1})$ for $\kappa>2$ is not integrable and approximated methods are needed to access its spectrum. 
SU($2$) and SU($\kappa$) fermions enjoy a different physics.
For spin one-half fermions, spin excitations, the so called spinons, are gapless in thermodynamic limit; charge excitations, instead, are gapped at half filling (Mott phase) and gapless otherwise \cite{andrei1995integrable}. In the low energy limit, spin and charge excitations separate each other.
Notably, the Mott phase is suppressed only exponentially for $\kappa=2$ \cite{lieb1968absence}. For $\kappa > 2 $, fermions display a Mott transition for a finite value of $U/J$ \cite{manmana2011n,cazalilla2014ultracold}. For incommensurate fillings, a superfluid behavior is found. In the SU($\kappa$) case, spin and charge excitations can be coupled \cite{affleck1988critical}.

%To close the section, we mention that relevant many-body quantum %systems can be realized by resorting to the notion of synthetic %dimension that is recently experimentally implemented by using cold %atoms quantum technology. The main idea expressed by the synthetic %dimension is to consider suitable degrees of freedom of the particles’ %system that can layer the dynamics of the system as a true spatial %dimensions \cite{PhysRevLett.108.133001,celi2014synthetic}. 
%Examples of synthetic dimension can be achieved through coupled %internal (hyper-fine) atomic states. The crucial requirement here is %that each internal state is coupled to only two other states in a %sequential way. For cold atoms, this can be achieved by suitable Raman %coupling between the different internal states. 
%This approach allows to study exotic physical conditions and %dimensionality effects beyond the spatial dimension of the matter-wave %circuit. 
%(even beyond three dimensions) with enhanced control and flexibility. %This way, many-body systems with exotic physical conditions have been %explored. 
%Examples include alkaline-earth atoms realizing bosonic ladders in %various conditions of gauge fields and spin-orbit couplings %\cite{PhysRevX.8.031045, buser2020interacting,livi2016synthetic} or %topological matter \cite{ozawa2019topological}.
%\note{WK: Replaced \cite{2018_marchukov_preprint} by %\cite{marchukov2019splitting}, please confirm. Luigi: Both Marchukov papers %have nothing to do with synthetic dimensions...}

\subsubsection{Impurities, weak-links and contacts}
Barriers, weak-links, quantum impurities and contacts are essential features for matter-wave circuits. Most, if not all, of these features can be experimentally realized, with a wide range of parameters both in the spatial and time domains. Below, we will sketch on how they can be incorporated in the systems Hamiltonian.

In continuous systems (like Eqs. (\ref{delta}), (\ref{GPE}), (\ref{GYS})), 
%ideal localized barriers can be inserted through delta-functions.
ideal localized barriers can be modeled as delta-function potentials. 
They can be used to stir ring-shape condensates \cite{hallwood_delta,hallwood2010robust,nunnenkamp2011superposition,schenke2011nonadiabatic}. In numerical simulations describing closely the experimental conditions
%and in experiments,
the delta function is replaced by a suitably peaked Gaussian function \cite{nunnenkamp2011superposition}.
Localized barriers in lattice systems are achieved through weak-links in the hopping amplitudes \cite{amico2014superfluid,aghamalyan2016atomtronic} or by suitable offsets of the local potentials \cite{aghamalyan2015coherent,cominotti2015scaling}.

In a typical transport set-up, the effect of a thin localized barrier of large strength can be described by the tunnel Hamiltonian
${\cal H}= {\cal H}_L+{\cal H}_R+{\cal H}_t$
in which ${\cal H}_L$ and ${\cal H}_R$ are the left and right leads, and ${\cal H}_t$ is the tunneling Hamiltonian. A standard expression for ${\cal H}_t$ is $H_t={\cal J} (\psi_L^\dagger \psi_R+ h.c.)$, $\psi_L$ and $\psi_R$ being single particle operators of the left and right leads respectively (see for example \cite{nazarov_blanter_2009}) and ${\cal J}$ the tunnel amplitude. At a semi-classical level, the two leads transport can be described in terms of the atoms transfer among the reservoirs $\Delta{N}=N_L-N_R$. Specifically, the current is defined as $I=-(1/2) d(\Delta{N})/dt$. Such a logic is applied in the two-terminals transport set-ups discussed in the Sect.\ref{Mesoscopic}. 
%In there, transport in open matter wave circuits is studied through standard %master equations \cite{breuer2002theory}. 

\subsection{Persistent currents in atomtronic circuits \label{sec:persistentCurrents}} 
%\onecolumngrid
%\begin{center}

%
Even though persistent currents are mesoscopic in nature, they are instrumental for atomtronics. They can provide an important tool for quantum simulation, since they can probe quantum phases of matter. At the same time, persistent currents can be used for atomtronic devices such as, e.g. quantum sensing (see Sect.\ref{sensors}) or neutral currents-based platforms for qubit implementations (see Sec.\ref{qubit}).
\subsubsection{The concept of persistent current}
\label{persistent}
The persistent current is one of the defining notions of mesoscopic physics \cite{buttiker1983josephson,Imry:1999aa,Imry:2002aa}: in an electronic ring-shaped gas (a metal for example) pierced by a static magnetic field, a dissipationless current can occur. This is a manifestation of the electron phase coherence all along the ring, implying that the coherence length is larger than the system size. This counter-intuitive phenomenon occurs in the quantum regime, when resistive effects due to interactions, presence of impurities and thermal fluctuations leading to decoherence are negligible. 
Persistent currents in electronic systems are thoroughly studied  both theoretically and experimentally (see eg \cite{zvyagin1995persistent,saminadayar2004equilibrium} and references therein), with the aim of shedding light on its own mechanism, studying the effect of interactions, understanding the role of the impurities \cite{matveev2002persistent,chakraborty1994electron,riedel1993mesoscopic,Imry:2002aa}. 

In superconductors and superfluids, the persistent currents coincide with the supercurrents
flowing across the ring and originate from the macroscopic phase coherence of such quantum 
states. Experimental observations of persistent currents are reported in several 
condensed-matter systems: normal metallic rings 
 \cite{levy1990magnetization,bleszynski2009persistent,mohanty1999persistent} and superconductors, \cite{deaver1991experimental}. 
Exciton polaritons have also been proposed as platform to study persistent currents under 
controllable conditions 
 \cite{sanvitto2010persistent,li2015stability,gallemi2018interaction,lukoshkin2018persistent}. 

By virtue of the control and flexibility of their operating conditions and the possibility to deal with different particles' statistics, ultracold atoms provide an ideal platform to study persistent current with a new scope. 
The study of persistent currents is first initiated  in cold atoms systems confined to ring-shaped potentials and pierced by a synthetic magnetic field by \onlinecite{amico2005quantum}.
Indeed, a quantum gas in ring-shaped confinement and subjected to an artificial gauge field with flux $\Phi$ (see Sect.\ref{hamiltonians}) behaves as a charged particle subjected to a magnetic field. The artificial magnetic field can be engineered by a variety of techniques in quantum technology ranging from a simple rotation to the transfer of angular momentum through two-photon Raman transitions or Berry phases and hologram phase imprinting techniques \cite{dalibard2011colloquium,goldman2014light-induced}. 
%\Luigin{Check references on artificial gauge field.} OK review by goldman added
The effective magnetic field imparts a phase gradient on the particles' wave function defining a finite velocity field along the ring. For sufficiently smooth guides (i.e. the most common situation in cold-atom experiments) the particles' flow is dissipationless. 
The current is obtained from the free energy thanks to a thermodynamic identity deduced from the Hellmann-Feynman theorem $I=-(1/2 \pi)\partial F/\partial \Phi$ \cite{zvyagin1995persistent}. In the
ground-state, the persistent current is $I=-(1/2 \pi) \partial E_{GS}/\partial \Phi$.

In the quantum-coherent regime, the particle current is predicted to be a periodic function of the applied flux $\Phi=\omega R^2$ of the artificial gauge field, $R$ being the ring radius.
A theorem due originally to Leggett \cite{leggett1991granular} shows that for spinless fermions and bosons with repulsive interactions on a clean ring, the persistent currents do not depend on the interaction strength, but merely reflect angular momentum conservation along the ring, so that the ground-state energy is written as $E=E_0 (\ell-\Phi/\Phi_0)^2$, with $\ell$ denoting the $z$-angular momentum quantum number, i.e. the ground-state energy is piece-wise parabolic and each parabola indicates a different value of angular momentum carried by the circulating particles.
The period of oscillation of the currents is the flux quantum $\Phi_0=\hbar/m$. Inclusion of localized impurities or of a barrier mixes the angular momentum states, thus smoothing the amplitude of the persistent currents \cite{hekking1997quantum,matveev2002persistent,cominotti2014optimal}. 
Such  an impurity is felt by the  interacting fluid as an  effective localized barrier affecting the system in a way that depends on interaction.
%\Luigin{Explain the interplay between barrier strength and interaction}. 
For repulsive interactions, the Luttinger liquid paradigm \cite{giamarchi2003quantum} holds at intermediate and strong interactions and the effective barrier depends on a power law with the ring size, while for the weak interactions, the barrier is screened by healing length effects \cite{cominotti2014optimal}. The regime where the barrier effectively splits the ring into two disconnected parts is a universal function of barrier and interaction strength \cite{aghamalyan2015coherent}.
For attractive interactions, the excitation spectrum is quadratic and a universal scaling with a non-trivial interplay of barrier and interaction strength is observed in some interaction regimes \cite{polo2020quantum}.

Relying on the enhanced capabilities of DMD's or painting techniques,  persistent currents can be engineered by machine learning assisted dynamics of the trapping potential\cite{haug2019engineering}. Specifically, the engineering can be  achieved by training a deep-learning network on the local potential off-sets thereby trapping the atoms in ring-shaped circuit with lumped parameters Eqs.~(\ref{BHM}), (\ref{Eq:QP}). This approach predicts  better performances in the state preparation and in the very nature of persistent currents (currents involving three angular momentum can be engineered) can be achieved, compared with the existing protocols based  stirring protocols. 

Persistent currents have been  also studied in bosonic ring ladders \cite{aghamalyan2013effective,haug2018mesoscopic,victorin2019nonclassical,polo2016transport,richaud2017quantum}. There, discrete vortex structures can occur in the transverse direction, giving rise to a wealth of phases (see eg for a review \cite{amico2021roadmap}). Josephson oscillations and orbital angular momentum dynamics in  two coupled rings in a stuck configuration have been studied in\cite{lesanovsky2007time,nicolau2020orbital,oliinyk2019tunneling}.
%\Luigin{also Veronica Ahufinger, review by GOldman-Dalibard, roadmap}
In multi-component mixtures, the criterion of stability of persistent current and the relation with entanglement  
are  addressed in  \onlinecite{anoshkin2012persistent,abad2014persistent,spehner2021persistent}. Transfer of angular momentum between different bosonic species was theoretically addressed in \onlinecite{penna2017two}

We close the section, by summarizing the important results based on Gross-Pitaevskii dynamics in two or three spatial dimensions. 
%\red{Luigi: Here we should review the papers of Charles Clark, Mark Edwards etc}
In most of the protocols studied so far, the matter-wave flow is obrained by  stirring.   
Many sources of decay of persistent currents have been identified: generation of elementary excitations, thermal fluctuations, vortices and vortex rings \cite{piazza2009vortex,mathey2014decay,wang2015resonant,abad2016persistent,xhani2020critical}. For a tightly confined toroidal shape condensate, persistent currents may still decay by phase slippage mechanisms, in particular through incoherent or coherent phase slips depending on interaction and temperature regimes \cite{danshita2012quantum,kunimi2017thermally,polo2019oscillations}. By this approach, stirring the matter wave is studied in in race-track atomtronic circuit\cite{eller2020producing}.

\subsubsection{Experimental observation  and read out of persistent current in bosonic toroidal-shape atomtronic circuits}
 
\begin{figure}[bt]
\centering
\includegraphics[width=0.5\textwidth]{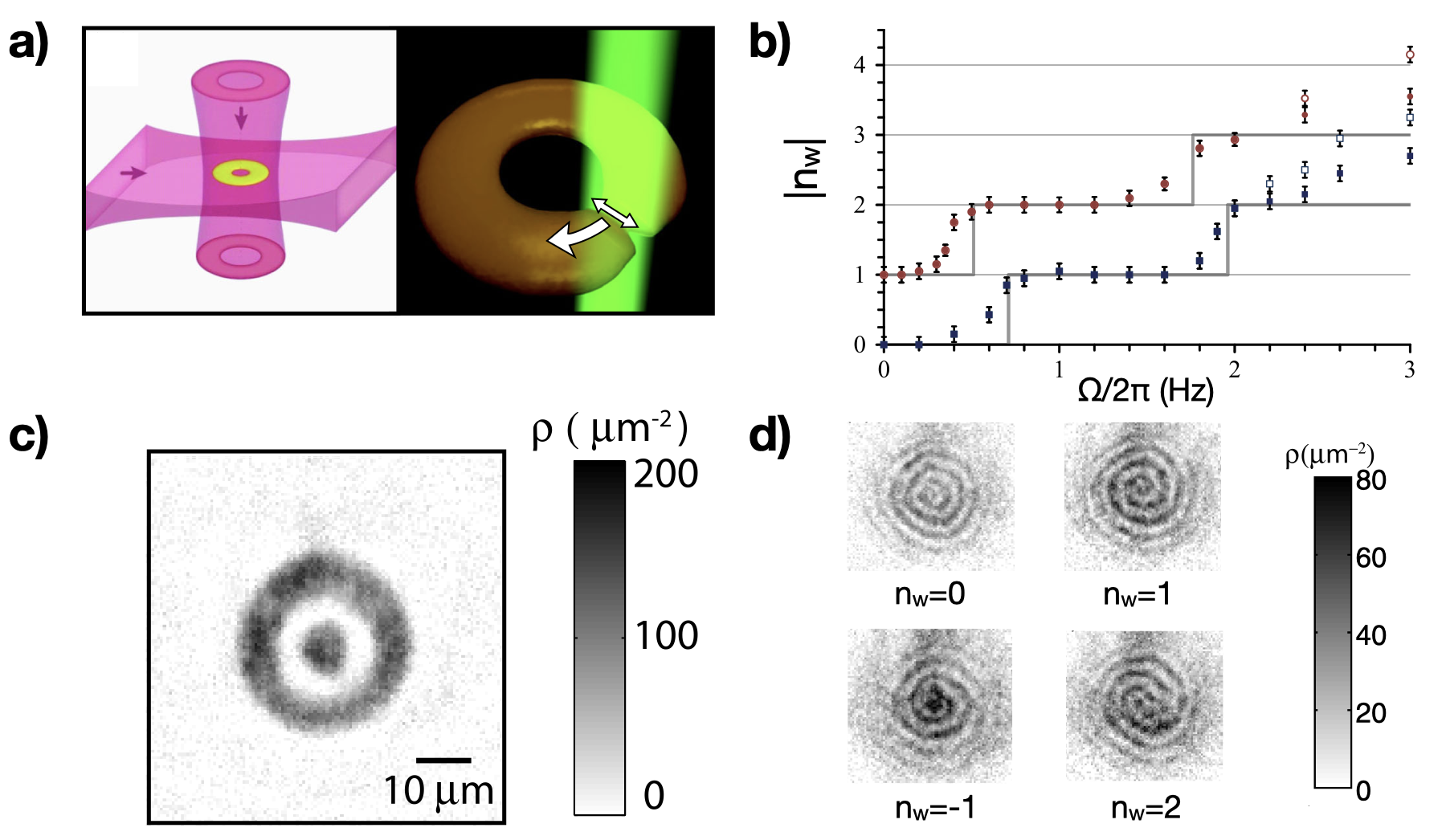}
\caption{
Experimental realization of BEC rotating in a circular atomtronic circuit. a) Schematics of the fabrication of the optical potential and the stirring protocol; the circular confinement is realized through a Laguerre-Gauss laser field; the transverse confinement is implemented through light sheets; the condensate is stirred with a rotating blue detuned focused laser beam. b)  The persistent currents  features quantized steps of the angular momentum imparted to the BEC expressed  by the winding number $n_w$. c) displays the ccd-contrast image  of ring shaped rotating BEC concentric with a second BEC; d) such a configuaration allows to probe direction and strength of the angular momentum through characteristic spiral interferograms.  Panels a) and b) adapted from \onlinecite{wright2013Driving}. Panels c) and d) adapted from \onlinecite{corman2014quench}.
%\note{WK: Reprint permissions for fig 5? Luigi: No since APS}
}
\label{fig:ExpCurrentRing}
\end{figure}
%
%\end{center}
%\twocolumngrid
Rotating fluids and persistent currents are observed in ultracold atomic gases on a ring in a donut-shaped ring trap \cite{ramanathan2011superflow,wright2013Driving, moulder2012quantized, ryu2014creation} - see Fig.\ref{fig:ExpCurrentRing}a) and Fig.\ref{fig:ExpCurrentRing}b).
A challenge to rotate a quantum fluid is the generation of excitations and vortices \cite{dubessy2012critical,arabahmadi2021universal}. The threshold for creation of excitations has been measured \cite{wright2013threshold}.
The decay of persistent currents due to thermal fluctuations has been also experimentally studied \cite{kumar2017temperature}.
Recent experiments has also achieved very high rotation quantum numbers, with a rotation speed up to 18 times the sound velocity \cite{pandey2019hypersonic,guo2020supersonic}.

By introducing two moving weak links on the ring in opposite directions, the transition from superfluid to resistive flow is studied \cite{jendrzejewski2014resistive}. This experiment provides a new technique: the use of a ring to address mesoscopic transport properties (see also Sec.\ref{Mesoscopic}). Along the same line, it is demonstrated that it is possible study the current-phase relation of a superfluid using a ring geometry \cite{eckel2014interferometric}.

Persistent currents can be explained with the various branches of the energy dispersion relation as a function of the flux or the rotation rate. In Ref. \cite{eckel2014hysteresis}, hysteresis among different branches is observed and proposed as a method to control an atomtronics device.
%Persistent currents are associated with various branches of the energy dispersion relation as a function of the flux. In Ref. \cite{eckel2014hysteresis}, hysteresis among the various branches is observed, and proposed as a method to control an atomtronics device.

\begin{figure}[!htb]
\centering
\includegraphics[width=0.5\textwidth]{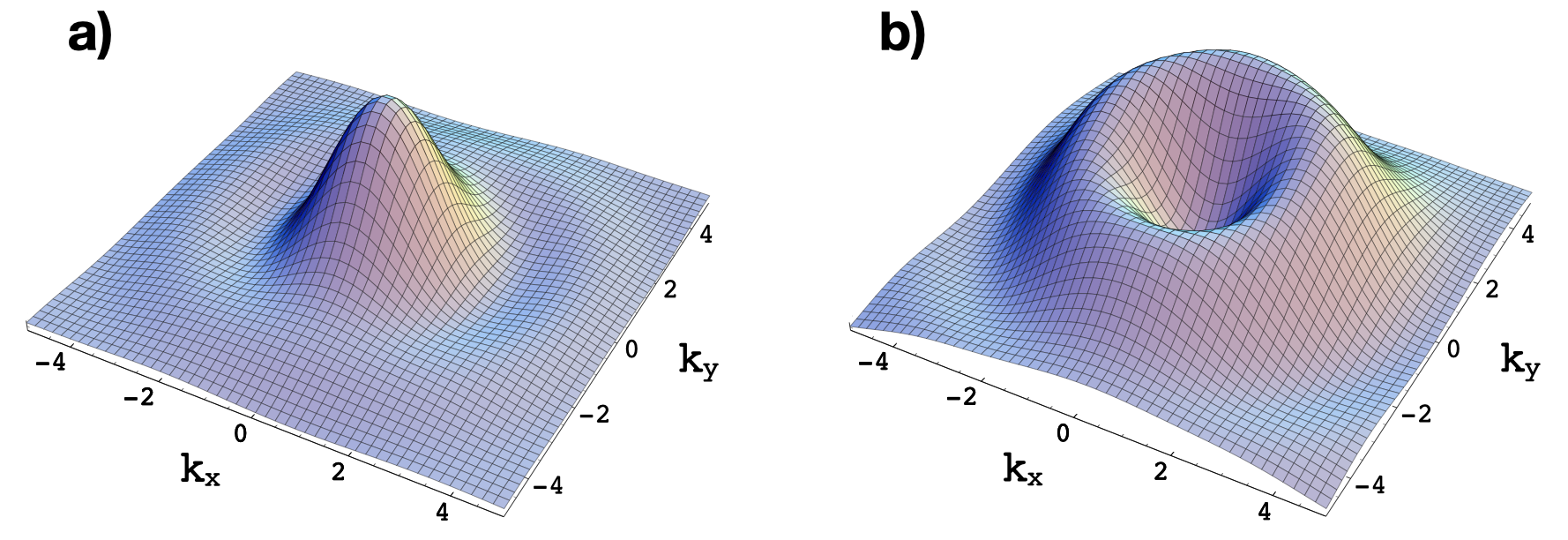}
	\caption{TOF expansion of a ring  shaped condensate pierced by an effective magnetic field. The figure display momentum distribution of a lattice system in the plane of the ring:   $n(\mathbf{k}) = |w(\mathbf {k})|^2\sum_{i,j} e^{-i (\mathbf{x}_i-\mathbf{x}_j) \cdot \mathbf{k}} \langle n \rangle$. The vector $\mathbf{k}=(k_x,k_y)$ and $w(\mathbf{k})$ are Fourier transform of Wannier functions.  a) and b) correspond to non-rotating and rotating condensate respectively. Adapted from \onlinecite{amico2005quantum} }
	%\note{WK: Reprint permission from journal for fig 6?; Luigi: No since APS}
	\label{fig:TOF_Amico}
\end{figure}

%\subsubsection{Readout of the current state \label{sec:Readout}}
Readout of the currents in ultracold atomic systems can be done in various ways. The ccd contrast image of the atoms density after long time releasing of the condensate  from the trap is called Time Of Flight (TOF).  In most of the experiments, the TOF image is achieved after $10-20 ms$ releasing time. The theoretical approach to TOF accounts to compute the momentum distribution of the system at the instant of time in which the trap is open $t=0$\cite{read2003free}. 
For condensate flowing along ring-shaped  circuits, TOF  displays a  characteristic shape in which the density around $\mathbf{k}$ is suppressed. The TOF image (taken from the top of the expanding condensate, along the falling direction) shows a donut shape which is very different from a bell-shaped image in absence of circulation\cite{amico2005quantum}, see Fig.\ref{fig:TOF_Amico}. The value of the donut radius results to change in discrete steps  corresponding to the quantization of angular momentum of the condensate \cite{moulder2012quantized, ryu2014creation,wright2013Driving,murray2013probing}- see Fig.\ref{fig:ExpCurrentRing}-b).

An important readout of the current state of the system is provided by heterodyne phase detection protocol \cite{corman2014quench,eckel2014interferometric,matthew2015self-heterodyne}. In this case, the ring condensate co-expands with a concentric disk condensate fixing a reference for the phase. The resulting image shows a characteristic spiral interferogram whose details (number of arms and sense of rotation) depend on the direction of circulation of the currents. The spiral interferograms are also sensitive to the possible phase fluctuations along the ring \cite{corman2014quench,roscilde2016from}: in this case, they display dislocations associated to phase slips - see Fig.\ref{fig:ExpCurrentRing}-b) and Fig.\ref{fig:ExpCurrentRing}-c). A minimally invasive technique based on Doppler shift of the phonon modes of the condensate has been demonstrated to be effective to measure the winding number \cite{Kumar_2016}.  

For correlated systems on lattices, the phase information can be achieved by studying noise correlations in the expanding density \cite{haug2018readout}.

\subsubsection{Persistent current in fermionic rings}

%\Luigin{Please check references. References to be added?} Cai2021 and Pecci2021 added

%Persistent currents occur both in bosonic and fermionic systems with %noticeable differences associated with the quantum statistics of the %particles and the interactions among its constituents. 
The first cold fermionic atoms  persistent current's analysis is carried out in \cite{amico2005quantum}. In this study, the fermions repel each other with an Hubbard interaction (see Eq.\ref{eq:1}). 
Persistent currents of Fermi particles are subjected to parity effects: the currents behave diamagnetically or paramagnetically depending on the parity of the number of particles on the ring \cite{leggett1991granular}.
%: starting from $\Phi=0$, at increasing flux, in the first case the current starts from zero and increases while in the second it starts from a finite value and it decreases. (Is this description needed?)
The effect is due to the periodic boundary conditions imposed on the wavefunction. Explicit calculation with bosonization shows subtle effects of interactions and the effect of temperature \cite{loss1992parity}. There is a critical temperature for the disappearance of the oscillations of the current as $k_{B}T=\hbar^2 mR^2$.
In the case of interacting bosons, no parity effect occurs. In this case, the response is always paramagnetic \cite{pecci2021probing}. %\note{WK please check reference}. 

Readout of current states by interferometric means for fermions requires some attention \cite{pecci2021phase} as compared to the bosonic case since all fermionic orbitals contribute to the interference pattern, giving rise to dislocations in the spiral interferograms. Also, the time-of-flight images of circulating current states display a visible hole only if the circulating current is large enough to displace the whole Fermi sphere in momentum space \cite{pecci2021phase}.

In the case when attractive interactions occur among the particles, pairing or formation of higher-order bound states (quartets, many-body bound states) directly affects the persistent currents \cite{byers1961theoretical}: the periodicity of the persistent currents scales as $\Phi_0/n$ where $n$ is the number of bound particles \cite{naldesi2022enhancing}. %\note{WK: Please check this reference. The cited article \cite{naldesi2020enhancing} is an updated version of the article 2019 } 
The curvature of the free energy at zero flux also displays a parity effect \cite{waintal2008persistent}: in this case, it arises from a new branch in the ground-state energy \cite{pecci2021probing}.  %\note{WK: please check updated reference \cite{pecci2021probing}}. 

Like the attractive bosons, the periodicity of the persistent current of repulsive fermions in the strongly correlated regime is reduced by $1/N$. This effect is demonstrated through Bethe Ansatz analysis for SU(2) \cite{yu1992persistent} and SU($\kappa$) Fermi gases \cite{chetcuti2022persistent}. Such behavior is due to the remarkable phenomenon of spinons-production in the ground state: spinons compensate the increasing effective flux; since the spinons are quantized and the magnetic flux changes continuously, the compensation can be only partial. Therefore, an energy oscillations with characteristic periods smaller than the bare flux quantum $\Phi_0$ is displayed.  Eventhough the  same $1/N$ reduction of the ground state periodicity is found in strongly correlated attracting bosons (occurring as result of formation of $N$-particle bound states in the `charge' quasi-momenta), here we note that the  'effective attraction from repulsion' resulting in $SU(\kappa)$ systems  arises because of the spin-spin correlations. \cite{naldesi2022enhancing}. Finite temperature  can affect the  periodicity of persistent current as result of interplay between thermal fluctuatuions and interaction\cite{pactu2022temperature}.

We finally note that, although the persistent current is mesoscopic in nature, it is demonstrated to display critical behavior when undergoing the quantum phase transition from superfluid to Mott phases that, for $\kappa>2$, occurs at a finite value of the interaction \cite{chetcuti2022persistent}. Mott transitions in multi-orbital $SU(\kappa)$ Hubbard models were investigated in \onlinecite{richaud2022mimicking,richaud2021interaction}. 
The first experimental demonstration of persistent current states in fermionic rings has been reported recently \cite{cai2022persistent,del2022imprinting}.
Focusing on  attractive interactions, both homodyne and heterodyne interference in the BEC regime have been obtained.   
An in-depth theoretical analysis of interference fringes of SU($\nu$) fermions was carried out in \cite{chetcuti2022interference}.

\subsection{Two terminal quantum transport in cold atom mesoscopic structures}
\label{Mesoscopic}

\begin{center}
\begin{figure}[ht]
\centering{
\includegraphics[width=0.5\textwidth]{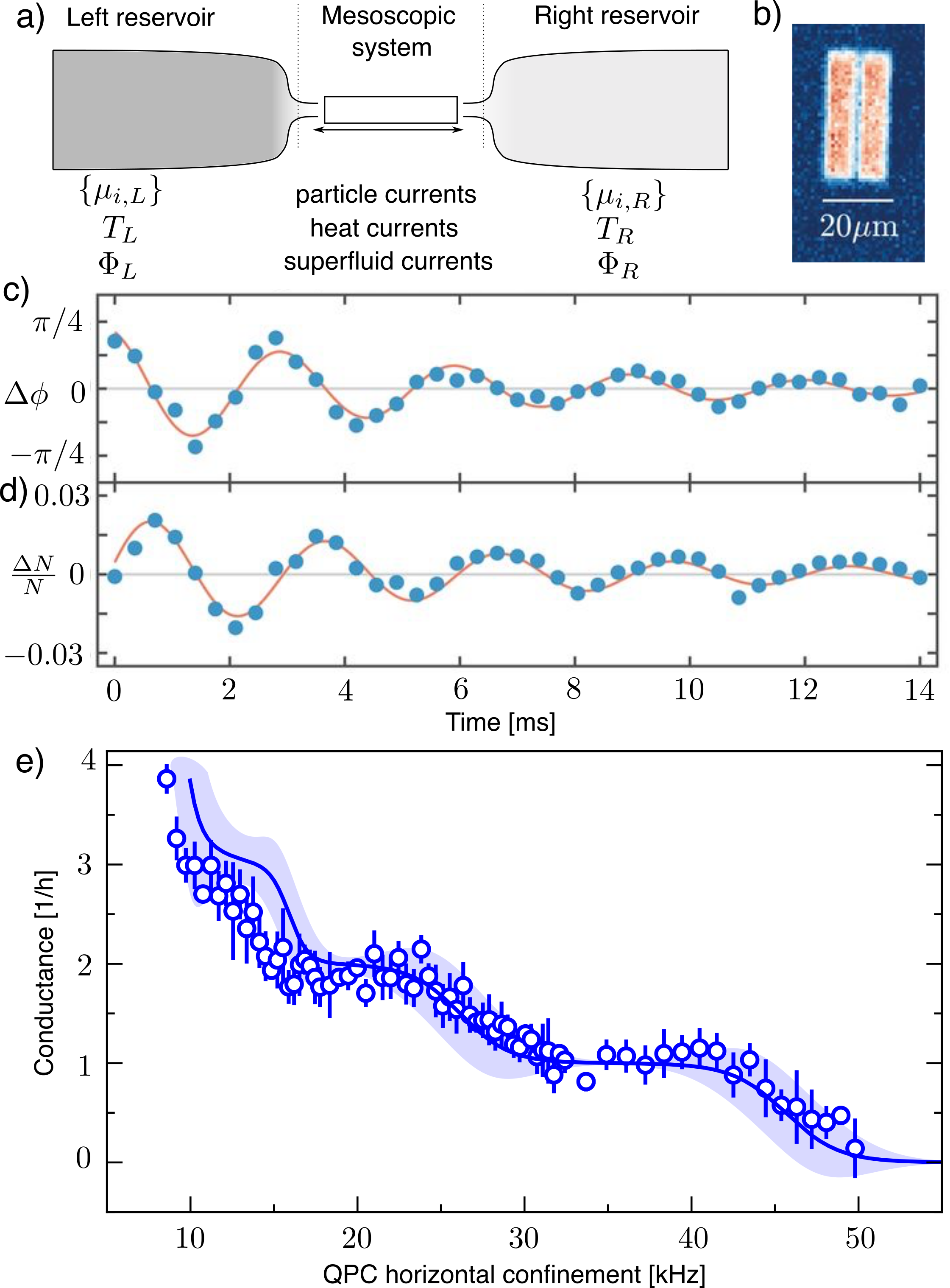}
	\caption{a) A two terminal system comprises a mesoscopic system connected to reservoirs ($L,R$), with a set of control parameters: chemical potentials for species $i$ $\mu_i$, temperatures $T$ or superfluid phase $\Phi$. b) Two-terminal Josephson junction in a strongly interacting two-dimensional Fermi gas. Phase difference c) and atom number difference d) as a function of time in the junction, after imprinting a relative phase difference of $\pi/4$. Adapted from \onlinecite{Luick:2020aa}\note{WK: Reprint permission from journal for fig 7?} e) Quantized conductance in a quantum point contact for weakly interacting Fermions. Conductance is measured as a function of the QPC trap frequency ($\nu_x$). Solid lines: predictions of the Landauer formula. Adapted from\onlinecite{Krinner:2015aa}}. 
	%Sketch of a Josephson junction consisting of two Fermi gases with chemical potential , particle numbers $N_{R,R}$, and phases $\phi_{L,R}$ separated by a tunneling barrier. (B) Absorption images of cold atom Josephson junctions. The width of the barrier is held fixed at a waist of $w=0.81(6)\,\mu$m while the size of the system is increased. (C and D) Time evolution of the phase difference (C) and relative particle number difference (D) between the left and right side of the box after imprinting a relative phase difference of $\pi/4$.  
	\label{fig:hamburgJosephson} }
\end{figure}
\end{center}

%The description in terms of lumped elements has a deep counterpart for the description of devices operating in the quantum mechanical regime: 
In a typical two terminals configuration, a \emph{mesoscopic} region, eg a channel or a ring, features quantum mechanical processes such as tunneling and interferences, and large leads characterized by their thermodynamic phases (which can be normal or superfluid) are connected to it to drive currents \cite{Imry:1999aa} - see Fig. \ref{fig:hamburgJosephson}a).

\subsubsection{Double well systems}
Two-terminal systems have been used with Bose-Einstein condensates (BEC) to observe and manipulate phase coherence \cite{andrews1997observation} (see \cite{dalfovo1999theory} for a detailed review). Conceptually, the simplest instance is a zero-temperature BEC in a double well potential, as originally proposed in \cite{smerzi1997quantum}. This system is of considerable interest from many perspectives, ranging from quantum metrology \cite{Pezze:2018aa} to quantum information processing \cite{Haroche:2006aa}. We restrict the discussion to atomic transport and refer the reader to these reviews for an in-depth discussions of these other aspects.

\paragraph{Tunnel regime}
In this regime, we focus on the dynamics of the population imbalance in the two wells \cite{smerzi1997quantum}, which is relevant for high barrier \cite{LeBlanc:2011aa,Spagnolli:2017aa}. In the unbiased, non-interacting regime, the dynamics reduces to Rabi oscillations of the population across the barrier at the tunnel period \cite{Spagnolli:2017aa}. For increasing interactions and weak population imbalance, Rabi oscillations smoothly evolve into plasma oscillations, with a frequency controlled by the repulsion between atoms \cite{albiez2005direct,levy2007the,LeBlanc:2011aa,pigneur2018relaxation}. For the largest imbalances, tunneling cannot compensate the effect of the non-linear interaction, leading to macroscopic quantum self-trapping \cite{albiez2005direct,levy2007the,Spagnolli:2017aa,pigneur2018relaxation}. This dynamics occurs in the absence of dissipation, which is true in the two-mode regime at zero temperature \cite{Gati:2006aa}. For attractive interactions, the plasma oscillation mode softens down to zero frequency at a critical attraction \cite{Trenkwalder:2016aa}.

The tunneling and interaction strength parameters of the two-mode model can be derived from the microscopic, mean-field Gross-Pitaevskii equation \cite{giovanazzi2008effective,LeBlanc:2011aa}, and depend on the details of the trap configuration. The predictions of this model are in good agreement with experiment \cite{ryu2013experimental}. The fluctuations due to the discrete nature of atoms allows us to describe quantum fluctuations of the phase, similar to phase noise in a non-linear interferometer \cite{Pezze:2018aa}. 
Beyond the mean-field approximation in the strongly interacting regime, a rich phenomenology has been predicted \cite{Zollner:2008aa}. %A promising experimental platform for investigating such physics has been demonstrated in \cite{Murmann:2015aa} with two atoms and strong correlations. 
Further quantum effects can also arise also from continuous quantum measurements of the atom numbers \cite{Uchino:2018ab}.

\paragraph{Extended reservoirs}
%When $\Delta V$ dominates the other energy scales, the dynamics is the simple Josephson oscillations encountered in the solid state context, with $\dot{\phi} = \Delta V $ and a current $N\dot{z}/2 = NK\sqrt{1-z^2} \sin\phi \sim NK \sin \Delta V t $.

%Review paragraph on the 'old observations' with optical barriers (Oberthaler, Steinhauer) and magnetic barriers (Treutlein, Schmiedmayer, Thywissen). 
While the two-mode approximation captures the essence of superfluid atomic currents through a tunnel junction, it disregards processes that take place within the reservoirs. Large reservoirs feature excitations that can couple to the current. The double-well structure is a very powerful configuration in which two identical systems can be produced and compared using interferometry. The internal dynamics of each system is then revealed in the phase relation between the two condensates. The latter has been used, in particular, for the study of one-dimensional gases in parallel wire configuration \cite{Hofferberth:2007ab,Hofferberth:2008aa,Betz:2011aa,Gring:2012aa,Langen:2015ab}. These landmark experiments reveal very fine details of the effective field-theory describing the one-dimensional reservoirs, including high-order correlations \cite{schweigler2021Decay}. Interestingly, allowing for a finite tunnel coupling between the two reservoirs modifies the effective Sine-Gordon model describing the low-energy physics \cite{gritsev2007linear}.
In head-to-tail geometry \cite{polo2018damping,tononi2020dephasing,binanti2021Dissipation}, corresponding to the realization of the Boundary Sine-Gordon model, the Josephson oscillations are damped by the phonon bath in each wire, realizing the Caldeira-Leggett model\cite{caldeira1983path}. 

%In addition, the two-mode model cannot capture the physics of superfluid Fermi gases as the Pauli principle forces atoms to populate a macroscopic number of modes. 

The Josephson dynamics coupled with that of the reservoirs is captured phenomenologically by the resistively-shunted Josephson junction model (RSJJ) \cite{tinkham2012introduction}, inspired from the condensed matter physics context and applied to atomtronics circuits in \cite{Eckel:2016aa,burchianti2018connecting,Luick:2020aa}. In this model, the reservoirs are described by an effective capacitance, corresponding to the compressibility of the gas $C = \frac{\partial N}{\partial \mu_r}$, with $\mu_r$ the reservoirs chemical potential and $N$ its atom number, derived from the equation of state and geometry. 

Large reservoirs are also described by a kinetic inductance due to the finite mass of the atoms, which adds to that of the junction to form the total inductance $L$. The frequency $\omega_0 = (L C)^{-1/2}$ represents the first normal mode of the system, reducing to the dipole mode of in a purely harmonic trap or to the plasma frequency in the two-modes model. The tunnel barrier itself is described by its critical current $I_c$, and the dissipative effects are captured by the parallel 'shunt' resistance $R$. The superfluid character of the system is encoded in the current-phase relation of the tunnel barrier $I=I_c \sin \phi$ and the Josephson-Anderson equation relating the chemical potential difference $\Delta \mu$ to the phase $\dot{\phi} = \Delta \mu$ \cite{Packard:1998aa}. %The RSJJ model reduces to the two-mode model in the plasma regime, where $R$ and $L$ can be neglected. From that point of view, the finite value of $L$ in the RSSJ corrects for the nonphysical divergence of the current predicted in the Rabi regime of the two-mode model with increasing tunneling rate. 
%The finite value of $R$ results in a dissipative, Ohmic decay of the atom number difference in the MQST regime, as the excess interaction energy is converted into heat. 

In this framework, the intrinsic properties of the superfluid junction can be studied independently of the dissipation by imposing a quasi-DC current \cite{levy2007the,Kwon:2020aa}. Alternatively, imprinting a phase difference across the junction by applying an external bias for a short time and measuring the current response through the junction realizes the equivalent of the DC Josephson effect \cite{Luick:2020aa}.

At non-zero temperature, thermally excited atoms serve as a natural source of dissipation justifying a finite value for $R$ \cite{Zapata:1998aa,Ruostekoski:1998aa,Marino:1999aa,levy2007the}. Even at zero temperature, a finite damping arises as the current couples to the internal dynamics of the reservoirs. In weakly interacting BECs and in Fermi gases in the BEC-BCS crossover, the physics captured by the resistance is related to the nucleation of topological defects \cite{wright2013Driving,jendrzejewski2014resistive,valtolina2015josephson,Eckel:2016aa,li:2016ux,burchianti2018connecting,xhani2020critical} or phase slips in one dimension \cite{polo2019oscillations, dubessy2021universal}.
The current also couples to Bogoliubov excitations in BECs and superfluid Fermi gases \cite{Singh:2020aa,Luick:2020aa}, competing with the coupling to vortices at higher temperature \cite{Singh:2020aa}. The coupling of tunneling with the Bogoliubov spectrum is theoretically studied in \cite{Meier:2001aa,uchino2020bosonic,Uchino:2020ab}, predicting a finite DC resistance at zero temperature.

Fermi superfluids also feature pair-breaking excitations, which have spectacular effects on the transport properties \cite{averin1995josephson}. Indirect evidences for such effects have been reported with cold Fermi gases in a point contact \cite{husmann2015connecting}. Remarkably, the Ambegaokar-Baratoff formula, relating pair-breaking excitations to the critical current in weakly interacting fermionic superfluids is shown to smoothly interpolate with dissipation induced by Bogoliubov excitations in the crossover from BCS to BEC \cite{Zaccanti:2019aa}.
Deeply within the scopes of atomtronics, the Josephson dynamics can be used to probe bulk quantum properties of the materials such as the 
 superfluid order parameter \cite{Kwon:2020aa} or flat-band superconductivity \cite{pyykkonen2021flat}.

\paragraph{Weak links in interacting systems}

In this regime, both the junction and the reservoirs have macroscopic size compared with the coherence or healing length of the gas. In such a weak link an appropriate description of transport can be obtained using superfluid hydrodynamics. It incorporates the transport of non-interacting thermal excitations \cite{Papoular:2014aa}. A lump element model can be derived from the microscopic hydrodynamics in a rigorous way for weakly interacting bosons, leading to accurate predictions for the dynamics of two-terminal systems \cite{Gauthier:2019aa}. In general, the population oscillations between reservoirs \cite{Papoular:2014aa} and that of the superfluid phase closely matches the plasma oscillation in the two-mode approximation. There, dissipation arises due to phase slippage mechanisms occurring within the weak link, and no qualitative difference emerges for long channels compared with tunnel-like barriers \cite{wright2013Driving,beattie2013persistent,jendrzejewski2014resistive,Eckel:2016aa,burchianti2018connecting,xhani2020critical}.% \Luigin{Refs on phase slips \Wolf{ \cite{wright2013Driving,beattie2013persistent}}

The case of superfluid Fermi gases has been studied in this context through a direct comparison between the unitary gas and a non-interacting Fermi gas, showing two-orders magnitude differences in the conductance \cite{stadler2012observing}. Furthermore, long channels can feature sections exposed to a tailored potential such as disorder. In a strongly interacting superfluid, a crossover is observed between a low disorder regime with superfluid transport and a disorder-dominated regime with low conductance \cite{Krinner:2013aa,Krinner:2015ac}. 

\subsubsection{Conductance measurements and incoherent reservoirs}

%\begin{center}
%\begin{figure}[ht]
%\centering{
%\includegraphics[width=0.4\textwidth]{quantized_freq.png}
%	\caption{Observation of quantized conductance in a quantum point contact imprinted onto a weakly interacting Fermi gas. Conductance is measured as a function of the QPC horizontal transverse oscillation frequency ($\nu_x$), for vertical oscillation frequencies of $10.9$ and $9.2$ kHz (blue and red, respectively). The solid lines show the predictions of the Landauer formula, without adjustable parameter. The shaded areas represent systematic uncertainties on the input parameters. The red curves are offset by two units vertically for clarity. Adapted from \onlinecite{Krinner:2015aa}}
%	\label{fig:quantized} \note{WK: Reprint permission from journal for fig 8?}}
%\end{figure}
%\end{center}

In situations where quantum coherence is either non-existent, or can be neglected such as in junctions dominated by dissipation, transport is captured by the conductance $G = I/\Delta \mu$, where $\Delta\mu$ is the chemical potential difference between the two reservoirs. 

\paragraph{Non-interacting atoms}
For reservoirs of non-interacting particles or quasi-particles, the current is determined by the energy-dependent transmission coefficients of the junction $\mathcal{T}_n$, where $n$ labels the transverse modes of the junctions, through the Landauer formalism \cite{cuevas2017molecular}
\begin{equation}
I = \frac 1 h \int d\epsilon \sum_n \mathcal{T}_n(\epsilon) \left( f_1(\epsilon) - f_2(\epsilon) \right)
\label{eq:landauer}
\end{equation}
where $f_1$ and $f_2$ are the energy distribution of particles in the two reservoirs. For Fermi gases and liquids, $f$ is the Fermi-Dirac distribution corresponding to the reservoir chemical potential $\mu_i$ and temperatures $T_i$. The formalism also applies to weakly interacting bosons above the critical temperature \cite{Nietner:2014aa,Papoular:2016aa,Kolovsky:2018aa}. %For charge-neutral systems of cold atoms, the chemical potential is only defined at equilibrium, implying in particular that current flow is much slower than the equilibration time. 
The strength of the Landauer paradigm is the separation between the quantum coherent part and the incoherent reservoirs, the latter featuring fast dissipation processes that are not described microscopically. The atomtronics Laudauer setup, with control over the reservoir properties, has motivated detailed theoretical studies of the dissipation dynamics through a comparison of the various microscopic descriptions \cite{Ivanov:2013aa,Chien:2014aa,Nietner:2014aa,Gallego-Marcos:2014aa,Kolovsky:2017ab}. 

The Landauer paradigm has been proposed for cold atoms in  \cite{Bruderer:2012aa,Gutman:2012aa}
and independently realized experimentally in \cite{Brantut:2012aa} using weakly interacting fermions in two reservoirs connected by a mesoscopic, quasi-two-dimensional constriction. For tight constrictions, the system behaves as a simple $RC$ circuit, with capacitors modeling the reservoirs, and the constriction the resistance. Measuring the decay constants of an initially prepared particle-number imbalance between the two reservoirs, and inferring the compressibility from the equation of state allowed to extract the conductance of the constriction. Early experiments focused on variations of conductance induced by changes of shape of the constriction or the introduction of disorder \cite{Brantut:2012aa}. In very recent experiments a similar system was used to investigate Anderson localization effects in two dimensions \cite{White:2019aa}. 

At zero temperature and low chemical potential difference, Eq. (\ref{eq:landauer}) yields $I=\Delta \mu/h j$, $j$ an integer. Each mode energetically accessible in the conductor contributes independently by $1/h$ to the conductance. In experiments, this is manifested in jumps of the conductance by $1/h$ as the Fermi energy reaches the successive transverse modes of the constriction, as observed in condensed matter devices \cite{Wharam:1988aa,Wees:1988aa} and in the atomtronics context \cite{Krinner:2015aa}, as shown in figure \ref{fig:hamburgJosephson}. 

On top of such an ideal one dimensional conductor, high-resolution optical methods allowed for the projection of structures as described in section \ref{sec:BarriersAndbeam splitters}, such as point-like scatterers. Measuring the conductance as a function of the scatterer's location produces a high resolution spatial map of the transport process, akin to scanning gate microscopy in the condensed matter context \cite{Hausler:2017ab}. Disposing several scatterers in a regular fashion produced a mesoscopic lattice that exhibits a band structure directly observed in the transport properties, demonstrating the ability to observe and control quantum interferences at the single scatterer level \cite{Lebrat:2018aa}.

The notions of reservoirs and channels can be interpreted in a more abstract way through the concept of synthetic dimension, using internal states of atoms \cite{celi2014synthetic} or vibrational states of traps \cite{Price:2017aa}. The two-terminal transport concept has also found a generalization through this mapping: a spin imbalanced Fermi gas provides a realization of two terminals in the spin space, and an impurity with engineered spin-changing collisions provide the counterpart of a point contact \cite{You:2019aa}. The use of vibrational states of reservoirs and constrictions as a synthetic dimensions then allows us to envision synthetic multi-terminal situations, where transport would be sensitive to chirality \cite{Salerno:2019aa}. 

\paragraph{Incoherent transport of interacting atoms}
The situation of quantum point contacts and one dimensional constrictions in the presence of interactions has been heavily investigated in the condensed matter physics context \cite{Imry:2002aa,cuevas2017molecular}. For two-terminal atomtronics systems this situation has been envisioned theoretically for bosons in \cite{Gutman:2012aa} with ideal reservoirs and in \cite{Simpson:2014aa} with superfluid reservoirs, in the framework of Luttinger liquid physics. For fermions, the point contacts and wires have been investigated experimentally in the deep superfluid regime for a unitary Fermi gas \cite{husmann2015connecting}, showing non-linear current-bias relations that could be traced back to multiple Andreev reflections \cite{krinner2017two-terminal}. This regime is expected to interpolate continuously with the Josephson regime as the transmission in the point contact is reduced \cite{averin1995josephson,Yao:2018aa}. This is further investigated by continuously increasing interactions from the free Fermi gas, showing quantized conductance, up to unitarity with non-linear response \cite{krinner2016mapping}. In the intermediate regime, the conductance plateau is observed to increase continuously from $1/h$ up to values as high as $4/h$ before disappearing close to unitarity, which could be either due to confinement induced pairing within the contact \cite{Kanasz-Nagy:2016ab,Liu:2017aa} or superfluid fluctuations in the reservoirs \cite{Uchino:2017aa}.

Transport in the one dimensional lattice, featuring a band structure, offers the possibility to explore the fate of metallic and insulating behavior as interactions are varied \cite{Lebrat:2018aa}. It is found that the band insulator evolves smoothly into a correlated insulator comprising of bound pairs with unit filling in the lattice, as interactions are increased 
%up to unitarity, 
providing evidence for the Luther-Emery phase \cite{giamarchi2003quantum}.

\paragraph{Spin and heat transport}

Transport in the two-terminal system can be generalized to spin in a two-component Fermi gas, where the total magnetization is conserved, and can be exchanged between two reservoirs. The linear response in currents is expressed through a matrix relating the currents of the two spin components to their respective chemical potential biases, with off-diagonal elements describing spin drag. In contrast with particle conductance, magnetization currents are very sensitive to interactions since collisions do not conserve the total spin current. Even in the absence of a constriction or channel, two clouds of opposite polarization relax very slowly to equilibrium especially at unitarity \cite{Sommer:2011uq}, where the spin diffusion coefficient saturates to a universal value. These experiments have been repeated for a metastable, strongly repulsive Fermi gas, providing evidence for a ferromagnetic instability \cite{Valtolina:2017aa}. In the case of a one dimensional quantum wire, the strongly attractive Fermi gas is found to behave as an ideal spin-insulator, as a consequence of pairing \cite{krinner2016mapping}. Another possibility to manipulate spin currents is created by the use of spin-dependent potentials, that are used to produce a spin valve from a quantum point contact \cite{Lebrat:2019aa}. 

Heat and energy transport can be investigated by introducing a temperature bias between the two reservoirs and observing energy flow through the channel. Heat and particle currents couple both through the thermodynamics of the reservoir due to finite dilation coefficients, and through genuine thermoelectric effect originating from the energy dependence of the transmission coefficient, as is observed in \cite{Brantut:2013aa,hausler2021interaction}. This also opens the perspective of Peltier cooling methods for quantum gases \cite{Grenier:2014aa,Grenier:aa,Sekera:2016aa}. In the case of the unitary Fermi gas, a similar experiment on quasi-one dimensional constrictions was performed, yielding a low heat conductance and a breakdown of the Wiedemann-Franz law, in qualitative agreement with theory \cite{pershoguba2019thermopower}, but a thermopower compatible with that of a non-interacting Fermi gas \cite{husmann2018breakdown}. Such a breakdown is also predicted for strongly interacting bosons within the Luttinger liquid framework \cite{Filippone:2016aa}.

\paragraph{Dissipative barriers}
As opposed to electrons, atomtronics devices allows for the engineering of atom losses. This has been investigated using electron microscopy with the creation of highly localized purely dissipative barriers \cite{Barontini:2013aa,Labouvie:2015aa}. The non-hermitian character of the resulting Hamiltonian supports the observation of coherent perfect absorption \cite{Mullers:2018aa}. Using an optical barrier involving spontaneous emission also produces dissipation in addition to the optical potential. This is studied in \cite{Corman:2019aa}. The interplay of these effects with interactions and fermionic superfluidity is investigated in \cite{damanet2019controlling}.

%\subsection{Extended and engineered structures}
%\subsubsection{Disorder}
%Anderson localization, other effects.
%\subsubsection{Scanning gates and point-like scatterers}
%Scanning gate microscopy, mesoscopic lattices, spin-dependent barriers, dissipating barriers.
%
%\subsection{Generalizations}
%Synthetic dimensions: Demler, Goldman
%Many-body effects, topology, multi-terminals, noise
%
\subsubsection{Two terminal transport through ring condensates}

Transport in circuits with closed architectures provides a direct way to explore the coherence of the system \cite{Imry:2002aa}.  At the same time, it provides an instance of integrated atomtronic circuits.
Consider particles injected from a source into ring-shape circuit pierced by an effective magnetic field, and collected in a drain lead. There, the phase of particles couples with the gauge field and transport displays characteristic Aharonov-Bohm interference patterns \cite{aharonov1959significance, AB_fundations_RMP,vaidman2012role,leggett1980macroscopic}, as studied in electronic systems \cite{lobos2008effects, marquardt2002aharonov,rincon2008spin,rincon2009features,shmakov2013aharonov,hod2006inelastic,jagla1993electron,gefen1984quantum,buttiker1984quantum,webb1985observation,nitzan2003electron}. 

Atomtronics allows the study of transport through ring-shaped circuits in new ways, with carriers of various statistics, tunable atom-atom interactions and lead-ring couplings \cite{haug2019aharonov,haug2019andreev}. 
%\note{WK: should we cite this \cite{haug2020quantum} %PhD thesis? It seems covered by the other two %references.}
Specifically, the non-equilibrium dynamics, described by Bose-Hubbard or discrete Gross-Pitaevskii models, is analyzed by quenching the particles spatial confinement in both closed and open configuration. 
Depending on the ring-lead coupling, interactions and particle statistics, the system displays qualitatively distinct non-equilibrium regimes with by different response of the interference pattern to the effective gauge field. In contrast to fermionic systems, the coherent transport of strongly interacting bosons does not display characteristic oscillations as function of the effective magnetic flux. A possible explanation for the suppression of the Aharonov-Bohm oscillations comes as a compensation between the phase of the condensate and Aharonov-Bohm phase. For a field theoretic explanation for the absence of Aharonov-Bohm interference in the circuit see \onlinecite{tokuno2008dynamics}.

%\Annan{Andreev reflections also mentioned before, possible to %merge?}
The transport through lead-ring interface can display a bosonic analogue of Andreev scattering: when a bosonic matterwave hits the lead-ring interface, it is transmitted to the ring with the emission of an matter-wave of negative amplitude, a 'hole', that is reflected backwards \cite{daley2008andreev,zapata2009andreev,watabe2008reflection}. Two-terminal transports through rings and Y-junctions are considered in \cite{haug2019andreev}. %The bosonic Andreev scattering occurs at the lead-ring interface and the absence of Aharonov-Bohm oscillations in the source-to-drain transmission in the strong lead-ring coupling regime is confirmed. 

A coherent transport can also be achieved through topological pumping, by driving periodically in time a system protected by a band gap \cite{thouless1982quantized,thouless1983quantization}. Such periodic drives are natural within atomtronics, thanks to the availability of re-configurable circuits \cite{mcgloin2003applications,gaunt2012robust,gauthier2016direct}. Topological pumping through source-ring-drain atomtronic circuit is addressed in \onlinecite{haug2019topological}. This way, topological bands and Aharonov-Bohm effect in interacting bosonic system are intertwined: the Aharonov-Bohm interference affects reflections by inducing specific transitions between topological bands. The system effectively works as a non-linear interferometer, in which the source-ring and the ring-drain act as beam-splitters. Interaction adjusts the transmission and reflection coefficients and entangles the propagating wave-functions in the two arms of the interferometer.

\section{Atomtronic Components and Applications}
\label{components}

In this section, we discuss the atomtronic circuital elements that have been considered in the literature. 
The first sections concern the atomic analogues of some circuit elements in  classical electronics. We conclude the section with atomtronic qubits inspired by quantum electronics.

\subsection{Matter wave optics in atomtronic circuits}
Transport in atomtronic circuits can be either coherent transmission like in photonic circuits or more like a superfluid similar to superconducting electronics.
A classical example is the decay of superfluid currents as described in Sec.\,\ref{sec:persistentCurrents}.
The main stumbling block in observing the coherent transmission of matter waves over macroscopic distances is the degree of roughness of the waveguides that are currently available. 

Until recently, except for straight guides formed by collimated laser beams, atomtronics is limited to the latter.
This situation changes recently with the first demonstration of coherent guiding over macroscopic distances in a ring-shaped matter wave guide \cite{pandey2019hypersonic}. 
It is now possible to (de)accelerate BECs in an optimal way to speeds of many times the critical superfluid velocity and an angular momentum exceeding $40000\,\hbar$ per atom without observable decay over time.
Matter-wave lensing or delta-kick cooling \cite{arnold2002adaptive, kovachy2015matter} has now been demonstrated using gravito-magnetic lenses inside of TAAP matter-wave guides, where BECs and thermal clouds have been collimated, thus reducing their expansion energies by a factor of 46 down to 800\,pK \cite{pandey2021atomtronic}. Delta-kick cooling with an optical potential is routinely used in waveguide atom interferometers to lower the energy of an expanded and collimated BEC below a few nK  \cite{krzyzanowska2022matter}. 

\subsection{Transistors, diodes, and batteries} 

The early work in atomtronic devices sought to emulate semiconductor material-based elements by considering neutral atoms in optical lattices, \cite{seaman2007atomtronics, pepino2009atomtronic, pepino2021advances} but work also sought simply functional duals by considering atoms confined to a small number of potential wells \cite{stickney2007transistorlike}. There are substantial differences in underlying physics, as well as practical differences, between these two approaches to device design.

Lattice based devices share clear analogies with electronic systems in periodic potentials characterized by band structure effects. At the same time, bosonic many-particles systems are unavoidably characterized by specific quantum correlations making their dynamics very distinct from the electronic one. 
Specifically, lattice-based atomtronic components deal with superfluids (instead of conductors) and heavily rely on the possibility to engineer a Mott insulating quantum phase that interacting bosons can undergo to, for integer filling fractions (number of bosons commensurate with the number of lattice points). Another effect without any classical electronics analog is the macroscopic quantum self trapping phenomenon that can hinder the transmission of a bosonic fluid through a potential barrier \cite{milburn1997quantum,smerzi1997quantum}. As a specific example of the semiconductor approach, an atomtronic diode can be conceived in analogy with the electronic P-N junction diode: the different concentration of electrons and holes in the P and N materials set a potential drop that can be modulated by an external voltage bias; the so-called forward (reverse) bias corresponds to a reduction of the potential drop for the electrons (holes) at the junction, and therefore a particles flow takes place. 
In the atomtronic diode, the junction is realized by facing commensurate and incommensurate lattices of condensates; an abrupt change of the chemical potential at the junction, playing the role of the voltage bias \cite{pepino2009atomtronic}.
This way, the control of the chemical potential can make the bosons from the commensurate to the incommensurate lattice of condensate, but not vice-versa.
The diode may be connected to two bosonic reservoirs kept at different chemical potential that play the role of the battery. Ultimately, the nonlinear device behavior arises from the {\it non linearity} of the atom-atom interactions that is a feature specific to the lattice systems.

%We have been nonchalant in referring to the battery in the diode context. It is %deeply significant to appreciate that 
%Here, we note that classical electronic circuits are non-thermal-equilibrium %systems whose dynamics is wholly driven by the presence of a battery (or other %source of electric potential), that is, the battery supplies power to run the %circuit. It is also significant to appreciate that a battery is fundamentally %associated with an internal resistance, which causes the battery to dissipate %energy as it also supplies power to a load. In atomtronic circuits, it is a BEC, %having finite chemical potential and temperature, that serves to provide chemical %potential and to drive the non-equilibrium dynamics of a circuit. And like the %electrical battery, a BEC-based battery providing atom current to a circuit will %exhibit an internal resistance. While the classical battery is always associated %with a positive resistance, a atomtronic (BEC) battery can exhibit either positive %or negative internal resistance, depending on whether the supplied current is %thermal or condensed, respectively \cite{zozulya2013principles}. An experimental %study of atomtronic batteries is carried out in
Here, we note that classical electronic circuits are indeed non-thermal-equilibrium systems whose dynamics is wholly driven by the presence of a battery (or other source of electric potential) supplying power to the circuit. It is also significant to appreciate that a battery is fundamentally associated with an internal resistance, which causes the battery to dissipate energy. In atomtronic circuits, it is a BEC with finite chemical potential and temperature, that serves to provide the 'bias'  driving the non-equilibrium dynamics of a circuit. And like the electrical battery, a BEC-based battery providing atom current to a circuit will exhibit an internal resistance. While the classical battery is always associated with a positive resistance, a atomtronic (BEC) battery can exhibit either positive or negative internal resistance, depending on whether the supplied current is thermal or condensed, respectively \cite{zozulya2013principles}. An experimental study of atomtronic batteries is carried out in\onlinecite{caliga2017experimental}.

\begin{figure}
 \centering
 \includegraphics[width=\columnwidth]{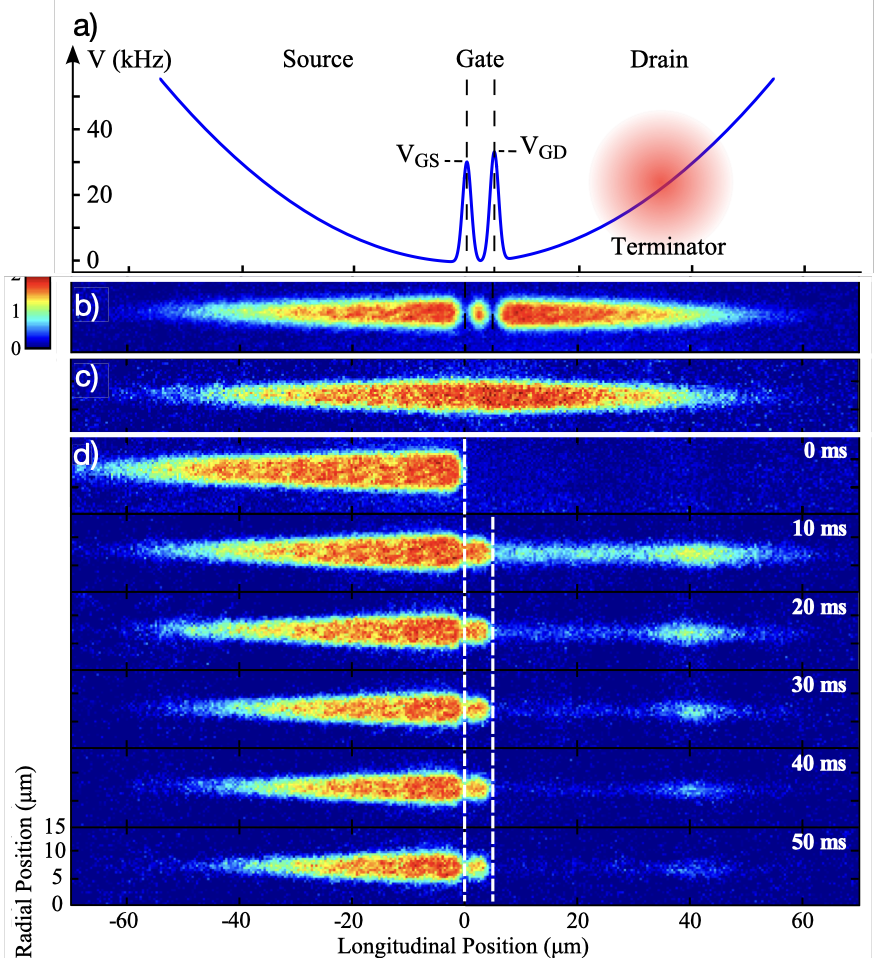}
 \caption{ The atomtronic triple-well transistor. The upper panel a) shows the atomic potential consisting of a hybrid magnetic confinement combined with barriers superimposed by optical projection of a pair of blue-detuned laser beams. A "Terminator" laser beam removes atoms from the drain by pumping them to an untrapped $m_F$ state. The gate width is $~4.5\mu$.  Thermal atoms are loaded in the hybrid potential (b) and in absence of the optical barriers (c). Panels d) and below are absorption images of the atoms in the wells at various evolution times starting from $t=0$ms and ending at $50$ms. 
 %Prior to $t=0$ a BEC is loaded into the source well, and the optical barriers are %loared to their specified heights to define $t=0$. These images are taken by %allowing the system to evolve for the specified time, but extinguishing the %terminator beam $4.8$ms before an absorption image is taken. In this time, atoms %  travel in the drain from the gate to the classical turning point at the %right-hand side of the drain potential. As the atoms are moving slowly as they %approach the turning point their density is high relative to other locations in %the drain. Note that the gate well is initially empty but in fact fills very %quickly. 
 Adapted from \onlinecite{caliga2016transport}.
% \note{WK: Reprint permission RECEIVED from Dana (220131-1700WK)}
 }
 \label{fig:transistor}
\end{figure}

A battery not only powers a circuit, but it is, of course, necessary to provide the gain associated transistor action. Such action has been studied in semi-classical context utilizing a triple-well atomtronic transistor in an oscillator configuration - See Fig.\,\ref{fig:transistor} \cite{stickney2007transistorlike,caliga2016principles}. 
The left-most well acts as source, the middle as gate, and the right as drain, where the nomenclature is taken from the electronic field-effect transistor. Here the system is initialized by placing a BEC at a given temperature and chemical potential in the source well. 

The transistor circuit behavior is characterized by a critical feedback parameter given by a normalized difference in barrier height \cite{caliga2012matterwave}:
$ %\begin{equation}
 \upsilon = \left ({V_{GD}-V_{GS}}\right )/({k_B T_S}) 
$,  %\end{equation}
in which $T_{S}$ is the temperature of the source atoms, and $V_{GD}, V_{GS}$ are the barrier heights (see Fig.\,\ref{fig:transistor}).  A semi-classical kinetic treatment has been developed in which the atoms are treated as particles, while they are also allowed to Bose-condense under appropriate conditions. With such an approach,  the BEC resulted to spontaneously develop in an initially empty gate well when the feedback exceeds a threshold value \cite{caliga2012matterwave}. This is reflected in the data of Fig.\,\ref{fig:transistor} where a high-density of atoms appears in the gate at $10$ ms evolution time --in fact the high density is apparent after only $1$ ms \cite{caliga2016transport}. 
The transport semiclassical dynamics of the transistor coupled to the environment, in which the atom steady currents are driven by the chemical potentials,   is studied in \onlinecite{caliga2016principles}. In particular, by analyzing the gain as function of the operating condition, it is proved that the such atomtronic component can be be used to supply power to a given load (therefore acting as an active component). 

% While the transistor experiments have been carried out with 

\subsection{The atomtronic quantum interference device}
\begin{figure}[htbp]
\centering
\includegraphics[width=0.47\textwidth]{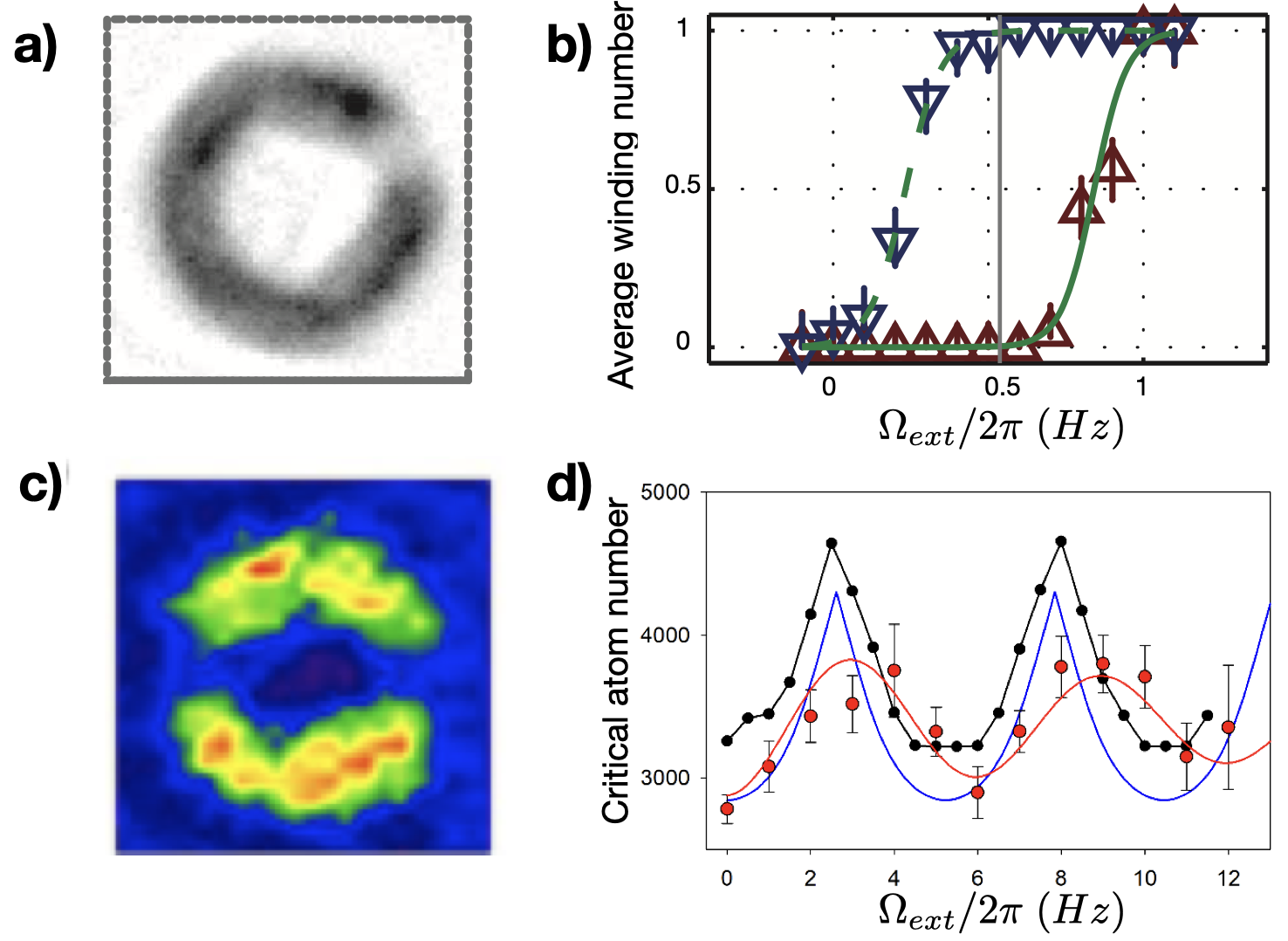}
\caption{Fabricated AQUIDs. a) The rf-AQUID of the NIST group. b) The hysteretic proprerty of the rf-AQUID. c) The dc-AQUID realized by the Los Alamos group. d) Oscillations of the critical current demonstrating superposition of superfluid currents. The black and blue curves are theoretical expectations and the error bars are result from the experiment, being the red curve a best fit of it. a) and b) adapted from \onlinecite{eckel2014hysteresis}; c) adapted from \onlinecite{ryu2013experimental} and d) adapted from \onlinecite{ryu2020quantum}.
\note{WK: Reprint permission from journal for fig 10?}}
\label{AQUID}
\end{figure}

A toroidal circuit of ultra-cold atoms interrupted by tunnel junctions provides the atomic counterpart of the SQUID: the Atomtronic Quantum Interference Device (AQUID). AQUID's with the characteristic control of noise and interactions and low decoherence of neutral ultracold matter, enclose a great potential both for basic science and technology. AQUID realized by a toroidal shaped superfluid Bose–Einstein condensate obstructed by a rotating weak link is carried out in NIST \cite{eckel2014hysteresis}, Fig.\ref{AQUID}a). By analogy with the radio frequency SQUID, such rf-AQUID displays hysteresis in angular momentum, Fig.\ref{AQUID}b). The role of the vortices generated by the stirring barrier \cite{yakimenko2015vortices,yakimenko2014generation} or thermal fluctuations \cite{kumar2017temperature,kunimi2019decay,mehdi2021superflow} have been analysed. Barrier strength and dynamical protocol of ramping up and down the stirring potential need to be carefully chosen to achieve a controlled and effective realization of the AQUID \cite{mathey2016realizing}.
The read-out of the rf-AQUID in the different regimes of interaction and barrier strength is studied in \onlinecite{haug2018readout} by monitoring the dynamics of interference fringes established after the condensate is released.   

Toroidal-shape condensates interrupted by two tunnel junctions have been experimentally fabricated by the Los Alamos group through the painting technique described in Sec.\ref{timeaveragedpotentials} \cite{ryu2013experimental}, Fig.\ref{AQUID}-c). Such system, providing the atomtronic counterpart of the direct-current SQUID, has dubbed as dc-AQUID. Following \cite{giovanazzi2000josephson} the dc Josephson effect in the experiment arises when the atom density (chemical potential) remain constant separately in each sector of the torus despite the two barriers move circumferentially toward each other. Indeed, the current increases
with barrier velocity until the critical current of the junctions is reached. At this point the system switches to the ac
Josephson regime characterized by an oscillating Josephson current. The frequency of the oscillations turns out proportional to
the chemical potential difference across the junction, but no net current across it. Remarkably, the critical current is observed to display characteristic oscillations demonstrating the superposition of superfluid currents, Fig.\ref{AQUID}d). 

The interference of persistent currents of dc-AQUIDs is recently carried out experimentally 
\onlinecite{ryu2020quantum}. 
By inducing a bias current in a rotating atomic ring interrupted by two weak links, the interference between the Josephson current with the current from the rotation creates a oscillation in the critical current with applied flux. This oscillation is measured experimentally in the transition from the DC to the AC Josephson effect. This experiment has been performed within a dilute Bose-Einstein condensate that is well described within a mean-field description and thus entanglement of currents, which is a key ingredient for the atomic qubit, has not been demonstrated. Nonetheless, it is a major step towards the implementation of the atomic qubit.

\subsection{Atomtronic qubit implementations}
\label{qubit}
Atomtronics qubit implementations have been proposed to combine the logic of cold atoms and superconducting circuits based qubits.
The basic idea is to use the persistent currents of cold atoms systems flowing in ring shaped potentials; in order to have two well defined energy levels, the translational invariance of the system needs to be broken by the insertion of suitable weak-links.
The presence of the weak-link breaks the axial rotational symmetry of the ring fluid and couples different angular momenta states, opening a gap at the degeneracy point among two angular momentum states (see Sec.\ref{persistent}).
This way, the two states of the qubit system are the symmetric and anti-symmetric combinations of the two angular momentum states~ \cite{amico2005quantum,solenov2010macroscopic,amico2014superfluid,aghamalyan2015coherent,aghamalyan2016atomtronic}.
The nature of the superposition state depends on the system parameters: at weak interactions it is a single-particle superposition, at intermediate interaction a NOON-like state and at very strong interactions a 'Moses state' i.e. a superposition of Fermi seas \cite{hallwood2006macroscopic,nunnenkamp2008generation,schenke2011nonadiabatic}.
\begin{figure}[htbp]
	\centering
\includegraphics[width=0.5\textwidth]{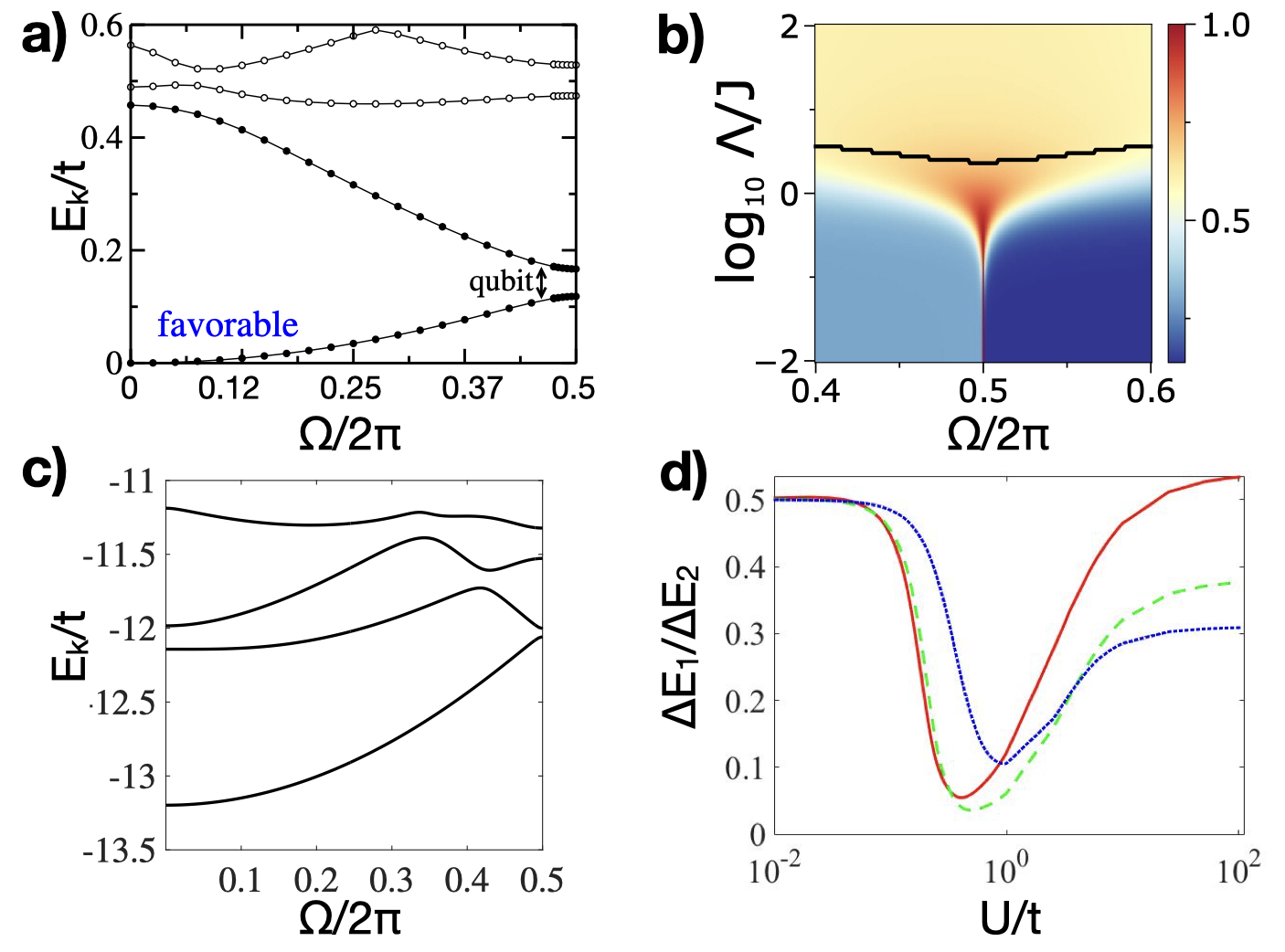}
	\caption{Atomtronic qubits. a) The upper panels refer to Bose-Hubbard rings interrupted by a single weak link. The bottom panels refer to a flux qubit configuration of a Bose-Hubbard ring interrupted by three weak links. a) and c) display the energy levels $E_k$. b) shows the noise correlations in the TOF image of the single weak link qubit. Panel d) summarizes  the  qubit quality factor as provided by the ratio between the energy gaps between the ground state energy and the first two excitation energies $\Delta E_1$ and $\Delta E_2$.  
	Panel a) adapted from \onlinecite{aghamalyan2015coherent}; panel b) adapted from \onlinecite{haug2018readout}; panels c) and d)  adapted from \onlinecite{aghamalyan_atomtronic_2016}.
%		\note{WK: Reprint permission from journal for fig 11? Luigi: No.  b) is % published in APS; the others are published in NJP authored by us}
		}
	\label{fig:Qubit}
\end{figure}
%Qubit dynamics exploiting the rf-AQUID %\cite{amico2014superfluid} or ring-shaped %condensates interrupted by three weak-links %\cite{aghamalyan2016atomtronic}, in analogy %of the Josephson junctions flux-qubits %configurations, have been demonstrated %theoretically integrating out the all the %phase differences amoung nearest-neighbour %condensates except the phase slip accross the %weak link, the effective dynamics results to %be governed by two level system Hamiltonian %with potential $V_f$. See %Fig.\cite{fig:Qubit}.
An important point in this context, is to establish to what extent the cold-atoms quantum technology would be capable to feasibly address the qubit. 
In particular, the energy gap separating the two energy levels of the qubit displays a specific dependence on the number of atoms in the ring network, atom-atom interaction and atom tunneling rates through the weak link \cite{nunnenkamp2011superposition}. 
%For the rf-AQUID qubit, the WKB gap separating the two levels of %the qubit is
%\onlinecite{aghamalyan2015atomtronics}
%\begin{equation}
%\Delta \simeq \frac{2 \sqrt{UE_J}}{\pi}\sqrt{ %(1-\frac{1}{\delta})} e^{-12\sqrt{ E_J/U} (1-1/\delta)^{3/2} }\,,
%\label{gap}
%\end{equation}
%
%in which $\delta \doteq {E_J}/{E_L} \ge 1$, with $E_L=J /M$, and %$E_J=J^{\prime}$. 
%
The numerical analysis based on BHM shows that the limit of a weak barrier and intermediate to strong interactions form the most favorable regime a qubit regime Fig.~(\ref{fig:Qubit}) \cite{amico2014superfluid,aghamalyan2016atomtronic}. 
The spectral quality of the qubit is analyzed in \cite{aghamalyan2015coherent} as function of the physical parameters of the system. The three weak links architecture \cite{aghamalyan2016atomtronic}, indeed, realizes a two-level effective dynamics in a considerably enlarged parameter space. 
Machine learning preparation of entangled persistent current is demonstrated in \onlinecite{haug2019engineering}.

Important for the aforementioned feasibility of the qubit dynamics is the analysis based on QPM working in the limit of  large number of particles  (see Eq. (\ref{Eq:QP})). Here, the  two level qubit dynamics emerges analytically\cite{amico2014superfluid}. For the case of a ring-circuit interruped by a single weak link, the effective Hamiltonian is $
\mathcal{H}_{\text{eff}} =
\mathcal{H}_{\text{syst}}+
\mathcal{H}_{\text{bath}} +  \mathcal{H}_{\text{syst-bath}}
$ 
in which  
$ %\begin{equation}
\mathcal{H}_{\text{syst}}=U  \bm{n}^2
+ E_L \bm{\varphi}^2
- E_J \ \cos(\bm{\theta}-\Omega)
%\label{jj}
$ %\end{equation}
where $ \theta$ is the phase slip across the weak link,  with  $E_L=J /M$, and $E_J=J^{\prime}$.
For $\delta \doteq {E_J}/{E_L} \ge 1 $, $\mathcal{H}_{\text{syst}}$ describes a particle in a double well potential.  $ \mathcal{H}_{\text{bath}}$ respectively 
describes the dissipative dynamics  $\mathcal{H}_{\text{syst-bath}}$ and  interaction due to the phase slips occurring in the other lattice sites. See also ~\cite{rastelli2013quantum}.

The qubit can be probed through a Rabi-type protocol: By quenching the effective magnetic field to the degeneracy point, characteristic Rabi oscillations occur with a frequecy $\propto 1/\Delta E_1$\cite{schenke2011nonadiabatic,polo2020quantum}.  
The two states of the qubit could be manipulated through a suitable 'pulse' of artificial magnetic field.

The read-out has been studied with various expanding condensate protocols \cite{aghamalyan2015atomtronics,haug2018readout}. 
%\note{WK: should we really be citing a PhD thesis %\cite{aghamalyan2015atomtronics} in an RMP? Especially, if it is not stored in %a permanent repository?}
In particular, the two-level system structure and the corresponding specific entanglement between the clockwise and anti-clockwise flows can be quantified through the noise in the momentum distribution:
$\avg{\op{n}(\vc{k})\op{n}(\vc{k})}-\avg{\op{n}(\vc{k})}\avg{\op{n}(\vc{k})}$, resulting to be  maximum at the degeneracy point - see Fig.\ref{fig:Qubit}b \cite{haug2018readout}.

%

%\begin{figure}[htbp]
%\centering
%\subfigimg[width=0.23\textwidth]{a}{TOFstdUWGL11N5s1p0%m0F0_5.pdf}
%\subfigimg[width=0.23\textwidth]{b}{TOFstdFWGL11N5s1p0%m0F0_4.pdf}
%\caption{Momentum noise $\sigma_k({\vc{k}=0})$ (in %color, normalized to one) plotted for potential %barrier $\Lambda$ against \idg{a} on-site interaction %$U$ (${\Omega=\frac{1}{2}}$) and \idg{b} flux $\Omega$ %(${U/J=1}$). Momentum noise is extracted from %time-of-flight image after long expansion. Only ring %is expanded, without central condensate. Black line %shows the critical point where depletion at the %potential barrier is 1\% of the average particle %number per site. Above the line the potential barrier %site is depleted. Other parameters are ${M=11}$ ring %sites and 5 particles. Figure %from~\onlinecite{haug2018readout}.}
%\label{CorrUvsp}
%\end{figure}

Proof of concepts for qubits coupling have been provided in which qubits are imagined to be arrange in stacks \cite{amico2014superfluid} or in planar \cite{safaei2018two} configuration. We comment that by relying on recent optical circuit design, much more flexible solutions are feasible to be implemented \cite{rubinsztein-dunlop2016roadmap}.

\subsection{Atomtronic interferometers}
 \label{sensors}
% \Wolfn{Maybe the title should read \emph{Quantum Sensors and Interferometers} in order to avoid a lonely subsection? Also,
% the sub-sub section headings don't make sense, e.g. they are all based on Sagnac..}
% \subsubsection{Interferometers}

% \Wolf{Something: The paths are the ones actually traveled by the atoms, which might be confined in two or three dimensions, i.e. freely travel in waveguides or be fully confined.}
An interferometer splits a wavefunction into a superposition of two parts and then recombines them in a phase-coherent fashion. 
If the wavepackets overlap perfectly at the output of the interferometer, the phase difference between the two arms is the difference between the phase shifts imposed by the pulsed beam-splitters and mirrors in each arm plus the propagation phase 	$\Delta \phi_{prop} = (S^1 - S^2)/\hbar$, where $S^i$ is the classical action computed along path $i$ \cite{peters2001high}. 
%erence in energy $(\Delta E)$ between the two paths leads over time to a phase shift $(\Delta \phi )$ between the two paths
$
% \begin{equation}\label{eq:PhaseOfIntererometer}
%	\Delta \phi = \frac{1}{\hbar}\int \Delta E(t) \, \diff t\,.
% \end{equation}
$
%At the second beam splitter, this phase difference is %translated into a population difference, which can %easily be read out. 
A beamsplitter at the interferometer transforms the phase difference into a population difference, which is easily read out.
%Light interferometry defines one of the most developed area of metrology for science and technology \cite{hariharan2010basics,padron2012interferometry}.
% Matter wave interferometers have been studied thoroughly on the past 50 years \cite{berman1997atom,cronin2009optics,barrett2014sagnac,kitching2011atomic,bongs2019taking}. 
Most of the atom interferometer solutions demonstrated to date involve free-falling atoms 
Traditional atom interferometers  involve free-falling atoms 
 \cite{stockton2011absolute,geiger2011Detecting,muller2008atom,sugarbaker2014atom,van2010bose,arimondo2009atom,bongs2019taking,geiger2020high}.
They  have the advantage of decoupling the atoms from many effects which otherwise might cause uncontrollable additional phase shifts, which can lead to a deterioration of contrast or a random shift of the fringes.
The main disadvantage is the size of the interferometer: Longer interrogation times lead to larger phase shifts.
Therefore free-falling high-precision matter-wave interferometers need to be very tall in order to accommodate the distance that the atoms fall during the interrogation---reaching a size of ten or even one hundred meters \cite{kovachy2015matter,Muntinga2013PRL}. 
In contrast to this, atomtronic interferometers use a trapping/guiding potential (usually magnetic or dipole) to compensate gravity and thus can achieve a much increased detection time with much reduced space requirements.
This comes, however, at the cost of an increased risk of noise and systematic effects due to fluctuations in the guiding potential.

\paragraph{Sagnac effect based atomtronic sensors: }
An important application of waveguide atom interferometer gyro-technology is inertial navigation in the absence of position information provided by a Global Navigation Satellite System (GNSS). An inertial navigation system (INS) contains three accelerometers, whose output is integrated twice to get displacement, along with three gyros that track the orientation of the accelerometers. It turns out that the navigation accuracy of current INS over timescales of hours and longer is limited by the drift in the zero of the gyros. These sensors are usually fiber-optic gyros (FOGs). Free-space atom interferometer gyros have already demonstrated extremely low drift \cite{gustavson1997precision,gustavson2000rotation,helm2015sagnac}, with their main disadvantage for some applications being the large physical size required to accommodate free fall of atoms being interrogated over several seconds. Guided atom interferometer gyros, analogous to the FOG, would be much more compact, making them attractive for navigation if they can be engineered to have low drift.

 In a typical rotation-sending configuration, atomtronic high-precision gyros are based on the Sagnac effect: two input quantum waves propagating along two different arms of a closed path circuit of enclosed area $A$ produce interference fringes at the interferometer output; if the circuit is rotated at rate $\Omega$, the interference fringes will be shifted by
\begin{equation}
\Phi_\mathrm{Sagnac}= {{4 \pi E}\over{hc^2}}\mathbf{A}\cdot\mathbf{\Omega}
\end{equation}
where $E$ is the energy of the traveling wave, and $\mathbf{A}$ and $\mathbf{\Omega}$ are respectively the enclosed area and the rotation vector. For frequency $\nu$ photon-based Sagnac interferometers $E_\mathrm{ph}= h \nu$ 
and for matter-waves it is $E_\mathrm{mw}=m c^2$ instead, yielding $\Phi_\mathrm{Sagnac}= {{4 \pi }\over{h/m }}\mathbf{A}\cdot\mathbf{\Omega}$. For equal particle flux and enclosed area, the difference in sensitivity between photon and matter wave interferometers is thus the ratio between the energies $E_\mathrm{mw}/E_\mathrm{ph}=10^{10}$. 
Light-based interferometers typically contain orders of magnitude more photons than the matter wave interferometers contain atoms. 
They also tend to enclose a much larger area.
Nevertheless, matter wave interferometers are expected to outperform their photon counterparts, e.g., where long-term stability is required.
% Of course, the final numbers must consider the constraints on the actual size the matter-wave interferometers can be realized and the matter-wave flux that can be made very favourable with photons. Despite the challenging in the actual technological implementations of matter-waves Sagnac interferometers, their use can provide an advantageous actual solution in specific situations. 
%\Annan{shall we cite \cite{pelegri2018quantum}: quantum device for measuring two-body interactions, scalar magnetic fields and rotations PROPOSAL... 
%\Wolf{I don't think that this really fits here, but maybe should be cited somewhere.}}
% Confined to a ring-shaped trap, a superposition of angular momentum modes of BECs could be used to measure two-body interactions, scalar magnetic fields and rotations \cite{pelegri2018quantum}.

\paragraph{Bright soliton rotation sensors: }  A BEC with attractive interactions (e.g. $^{85}Rb$ or $^7Li$) in a ring shaped guide can realize bright soliton interferometry: A localized barrier can split the solitons into two waves propagating in clockwise and anticlockwise directions that can ultimately recombine after traveling two semicircles. Even though (perfect) bright solitons can go through each other without changing their density profiles, the two waves can provide a Sagnac phase shift \cite{mcdonald2014bright,helm2012bright,helm2015sagnac,polo2013soliton}. The splitting  of  bright solitons scattering on a localized barrier is analyzed in \cite{helm2014splitting,weiss2009creation,marchukov2019splitting}. In such process, superposition states are predicted to occur \cite{streltsov2009scattering}. 
The roles of both quantum noise and interactions for rotation sensing with bright solitons described by a many-body Schrödinger equation have been analyzed by a variational principle \cite{haine2018quantum}. Because of the formation of solitons, enhanced control on the number of atoms $N$ in the experiments can be reached, that is expected to be beneficial for the sensitivity of the interferometry.  

The equivalent of a bright soliton in the fully quantum regime of a ring lattice of attracting bosons described by Bose-Hubbard model was studied in \onlinecite{naldesi2019rise}. Because of the lattice, the soliton and the number of atoms are protected by a finite gap. A barrier can split such 'quantum soliton' depending on the interplay between interaction, number of particles, and barrier strength. For ring-shaped confinement, it is demonstrated that the elementary flux quantum is reduced by $1/N$, where $N$ is the number of particles \cite{naldesi2022enhancing}. Such an effect potentially yields a $N$-factor enhancement in the sensitivity of attracting bosons to an external field that can reach the Heisenberg limit \cite{naldesi2022enhancing,polo2020quantum}.
%
%%% this is the Ron Folman paper
\begin{figure}
\begin{tikzpicture}
\node[inner sep=0pt] at (0,1.2) {     \includegraphics[width=0.48\textwidth]{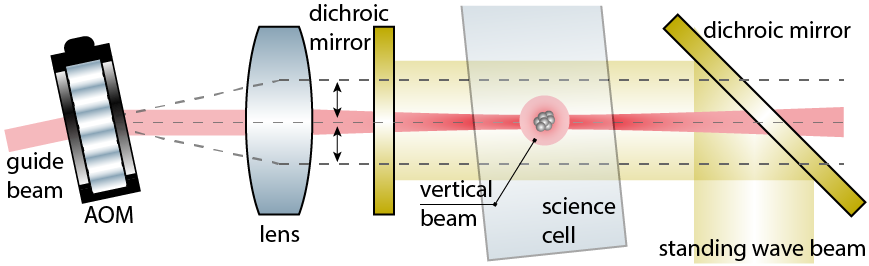}     };
\node[inner sep=0pt] at (0,-1.8) {     \includegraphics[width=0.48\textwidth]{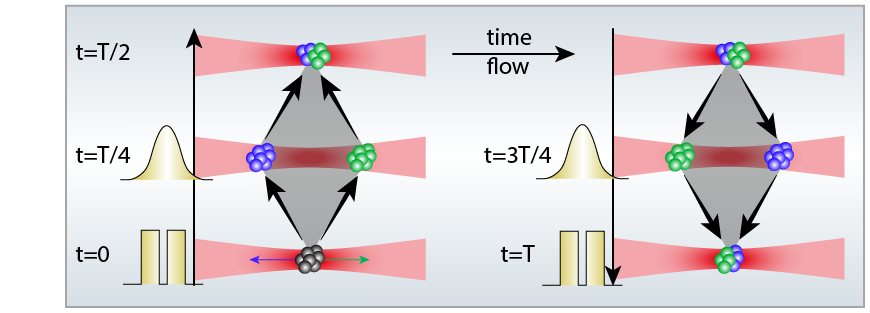}     };
\node[ align=left, rotate=-90] at (0.28,-4.3) {     \includegraphics[height=7.93cm]{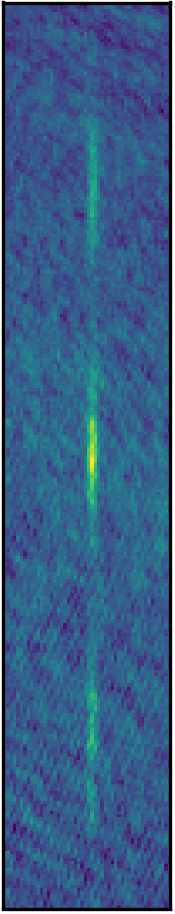}     };
\node[ align=left] at (-4.0, 2.2)     {a)};
\node[ align=left] at (-4.0, -0.6)     {b)};
\node[ align=left] at (-4.0, -3.7)     {c)};
\node[ align=left] at (2.7, -5.25)     {$p=2 \hbar k$};
\node[ align=left] at (0.4, -5.25)     {$p=0$};
\node[ align=left] at (-1.9, -5.25)     {$p= -2 \hbar k$};
\end{tikzpicture}
\caption{\label{fig:ExpScheme} a) Experimental setup. b) Moving waveguide interferometer. Blue and green colors of atoms correspond to the $+ 2 \hbar k$ and $-2 \hbar k$ momentum components respectively. The time flow follows the black arrows. c) Experimental data acquired $\Delta t = 12msec$   after the recombination pulse, with atoms in two channels:  $|p=0\rangle$ and $|p=\pm 2 \hbar k\rangle$. Reproduced from \onlinecite{krzyzanowska2022matter}}
\end{figure}
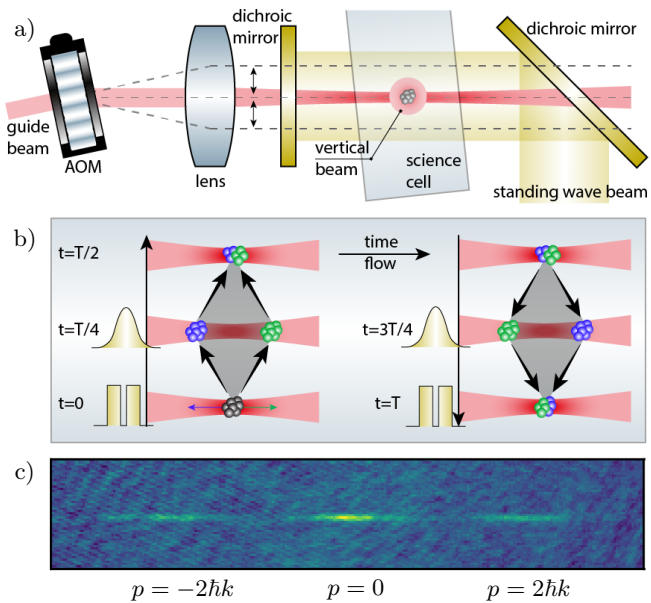
\paragraph{Demonstrated atomtronic interferometers: }
The first compact atom interferometers utilized stationary clouds of ultracold atoms. These devices and some of the notable physics resulting from experiments with them are discussed in \cite{schumm2005matter,gunther2007atom,jo2007phase,bohi2009coherent,riedel2010atom}. More recently, atomtronic interferometers with moving atoms have been realized in both optical and magnetic traps.

%%%%% waveguides i.e. untrapped in one direction
An early example is a Michelson interferometer using a BEC propagating over 120\,$\mu$m in a magnetic waveguide on an atom chip \cite{wang2005atom}. 
Smoother waveguides obtained with larger coils have been used to realize atom interferometers with thermal atoms \cite{wu2007demonstration, qi2017magnetically} and BECs \cite{burke2008confinement,burke2009scalable,garcia2006bose}. 
This approach has been used to measure the ground state polarizability of $^{87}$Rb \cite{deissler2008measurement}. 
In optical waveguides \cite{ryu2015integrated,akatsuka2017optically} linear interferometers extending up to 1\,mm have been demonstrated \cite{mcdonald201380hk}.
%
% \paragraph{Area enclosing interferometers}
%
A number of area enclosing interferometers have been realized in macroscopic magnetic traps \cite{wu2007demonstration,burke2009scalable, qi2017magnetically,moan2020quantum}. 

Recently, an atomtronic Sagnac rotation sensor based on a moving linear waveguide formed by a collimated laser beam is demonstrated \cite{krzyzanowska2022matter}. The 3.5\,mm$^2$ enclosed by the atomtronic circuit is the largest value realized to date. 

In area-enclosing waveguide atom interferometers the signal can be increased by allowing the wavepackets to make multiple orbits around the waveguide loop to increase the enclosed area. The maximum number of round trips is usually limited by atom loss when the counter-propagating wavepackets move through each other. It has recently been shown that this limitation can be removed in an interferometer based on a non-interacting $^{39}$K BEC, allowing for over 200 round trips in the guide \cite{kim2022one}. 

It is interesting to note that an atomtronic interferometer can be based on free propagation in a guide \cite{akatsuka2017optically,wang2005atom} or or on moving fully-trapped atom clouds (clock type interferometers) \cite{stevenson2015sagnac,navez2016matter}
% \Luigin{I suggest Wolf expands on his \cite{navez2016matter}. 
The first case can be pictured as the atoms functioning as an inertial reference much like a flywheel. 
The phase shift occurred by the fully trapped matter waves is perhaps best understood as being based on the relativistic time gains of an atom clock \cite{hafele1972around-the-world}. 

%\paragraph{Other atomtronic rotation sensors: }In %\cite{japha2007using}, the analogy between the Sagnac effect for %massive neutral particles and the %Aharonov-Bohm effect in %coherent electron transmission through mesoscopic rings is %explicitly employed to propose a new type of rotation-sensitive %guided atom interferometer. 
Finally, we note that several other schemes  for novel types of atomtronic interferometers have been proposed 
\cite{pelegri2018quantum,helm2108spin,helm2015sagnac,japha2007using,halkyard2010rotational,moukouri2021multipass,marti2015collective}.

\section{Remarks and Future Perspectives} 
\label{future}
%\Wolfn{Luigi -- please add %some info on 3D}
%\Luigin{The following red paragraph is out of place}
% \red{\emph{Quantum transport} In parallel with the %development of current-carrying devices and %structures, which has been the focus of this review, %the study of transport in 'bulk' quantum gases has %also seen tremendous progress in the last decade. A %detailed description of these aspects is beyond the %scope of this review, but the combination of these %studies with microstructures and devices opens very %exciting perspectives. The transport of quantum %information and spread of correlations has been %investigated in the framework of many-body %localization \cite{abanin2019colloquium}, where %disorder yields a breakdown of thermalization and %entanglement spreading between parts of the system. %The ability to control the spreading of quantum %information by applying disorder opens the %perspective of atomtronic devices with controlled %thermalization rates. In optical lattices, transport %is a testing bed for the understanding of the physics %of the Hubbard model, in particular for Fermi gases %where classical simulations are notoriously difficult % \cite{anderson2019conductivity,brown2019bad,nichols20%19spin}. }

%Research in Atomtronics is entering a very exciting and interesting stage with a substantial impact on both basic and applied science. 

Atomtronics defines  micrometer-scale coherent networks to address both technology and basic science. It combines bottom-up and a top-down approaches: 
On one hand, the circuit elements can be designed to implement the microscopic theory in an experimental realization of unprecedented precision. Then, just like in electronics, different circuit elements can be assembled using a hierarchy of heuristic principles.
On the other hand, a circuit or even a single circuital element in its own can be used as a current-based quantum simulator to probe the correlated matter. 
%The discussion above has shown that the current degree of control and flexibility reached by the cold atoms quantum technology makes atomtronic circuits ideal platforms to study fundamental aspects of quantum coherent systems. %This way, 

Important domains of quantum many-body physics in restricted geometries, ranging from intermediate to extended spatial scales, now become accessible:
%can be explored:
Analogous to the analysis of current-voltage characteristics in solid state physics, atomtronic circuits have the potential to define current-based emulators and simulators, effectively widening the scope of the existing ones. 
% \cite{dowling2003quantum,bloch2005ultracold,buluta200%9quantum,cirac2012goals,lamata2014epj}. 
 Currents in particular are the natural quantity to explore not only superflows but also transport in disordered and complex media as well as topological properties and edge states. 
An interesting direction to go is to exploit atomtronic circuits to address important questions of high energy physics, such as the phase diagram of quark-gluon plasma \cite{cazalilla2009ultracold,he2006superfluidity,rapp2007color,ozawa2010population,chetcuti2021probe} or various scattering process in elementary particles physics \cite{fu2020jet,clark2017collective,surace2021scattering}. 
%\LeongChuan{A ring trap is analogous to a microscopic particle %collider. Ref: Fu, H., Zhang, Z., Yao, K. X., Feng, L., Yoo, J., %Clark, L. W., ... & Chin, C. (2020). Jet Substructure in %Fireworks Emission from Nonuniform and Rotating Bose-Einstein %Condensates. Physical Review Letters, 125(18), 183003, and Clark, %L. W., Gaj, A., Feng, L., & Chin, C. (2017). Collective emission %of matter-wave jets from driven Bose–Einstein condensates. %Nature, 551(7680), 356-359.} 
Bosonic rings can be employed to study the dynamics of the expanding universe \cite{eckel2018rapidly}.

Atomtronic circuitry has a practical potential as well — a potential that can be realized in part by leveraging the know-how and heuristic design principles of electronics. Atomtronic triple-well transistors are in many respects close analogs of their electronic field-effect transistor counterparts and can be utilized in matter wave oscillators, for example, to produce matter waves with high spatial coherence \cite{anderson2021matterwaves}, which in turn can carry modulated signals or be used in sensing applications. In the future, one can expect many of the familiar elemental functions of electronic circuitry such as amplifiers, switches, oscillators, and so forth, to be carried over to the quantum regime. In another direction, coupled ring circuits, ring-rectilinear wave guides etc have been considered as simple instances of integrated atomtronic circuits\cite{ryu2015integrated,safaei2019monitoring,perez2021coherent,polo2016geometrically}

%\Wolfn{'coherent matter waves' needs more explaining. Maybe write 'a coherent state of matter waves'?}
%Atomtronic devices have become the work-horses of many ultra-cold atom experiments. Arguably, they are by now the most reliable and certainly most compact source of coherent neutral matter-waves. 

Building on the theoretically demonstrated qubit dynamics of specific matter wave circuits (see Sect.\ref{qubit}), it will be certainly important to explore atomtronics as a platform for quantum gates. At the same time, matter-wave circuits provide a valuable route to realize high precision compact interferometers working on a wide range of sensitivity and very controllable physical conditions. Such devices are of considerable technological importance in different contexts ranging from inertial navigation \cite{bongs2019taking} to geophysics \cite{jaroszewicz2016review}.
Unlike their classical or quantum electronic counterparts, atomtronic circuits can operate a regime in which quantum effects can be dominant and long coherence times are possible with a much simpler cryogenics. 
In this context, experimental, theoretical and technological inputs are envisaged to be combined together to realize the optimal building block circuit from which complex structures forming actual devices and sensors can be constructed. 
An important challenge to face in the years to come is to integrate the atomtronic circuits with other existing technologies such as photonic or superconducting integrated circuits\cite{PhysRevLett.101.183006,mukai2007persistent,nirrengarten2006realization,muller2010trapping,tosto2019optically,muller2010programmable,PhysRevA.85.013404}(for hybrid circuits specifically  relevant for quantum information, see \cite{xiang_hybrid_2013,yu2016charge,yu2017superconducting,yu2016superconducting,yu2016quantum,yu2018stabilizing,yu2017theoretical,yu2018charge,verdu2009strong,petrosyan2019microwave,bernon2013manipulation,hattermann2017coupling, fortagh2007magnetic}).
Such hybrid networks may provide a valuable route for the fabrication of integrated 3D matter-wave circuits in which rectilinear, ring guides, beam splitters etc, together with the fields for the control and read-out of the quantum states and the lasers needed for cooling and manipulation of the cold atoms are built into a single chip. Such an approach can be important to achieve scalable matter-wave circuits. 

Both for studies in fundamental science and circuit design with wider specifications, an interesting future direction is to expand the investigations to fermionic atomtronic circuits\cite{cai2022persistent,del2022imprinting} or to open the research in the field to new platforms, as e.g.\,fermionic systems with $N$ spin components\cite{chetcuti2022persistent,chetcuti2021probe,chetcuti2022interference} and Rydberg atoms. In the latter platform, bath engineering \cite{uchino2020bosonic, keck2018persistent, damanet2019reservoir,damanet2019controlling} together with the achieved control of the Rydberg blockade phenomenon \cite{valado2016experimental,simonelli2017deexcitation,archimi2019measurements} can be explored to start currents with novel specifications. Such a solution may grant the access to the realization of fast atomtronic circuits.

\begin{acknowledgments}
We acknowledge the late Frank Hekking for the important contribution  given to the Atomtronics field from the earliest stages.
We thank G. Birkl, P. Bouyer, W.J. Chetcuti, C. Clark, J. Dalibard, R. Dumke, R. Folman, B. Garraway, T. Giamarchi, T. Haug, T. Neely, S. Pandey, H. Perrin, J. Polo-Gomez, C. Sackett, and J. Schmiedmayer for comments and suggestions on the manuscript. We acknowledge fruitful discussions with J.I. Latorre, A.J. Leggett, C. Miniatura, P. Naldesi, G. Pecci, and K. Wright.
\end{acknowledgments}

%\section{Conclusions} (AMICO, MINGUZZI)
%\label{conclu}
%\input{conclusions}
%\newpage

%There are all the conditions for which the research in Atomtronics will be boosted in the next years. A comprehensive and well done review would be a very valuable tool for the broad quantum technology community. 

%\bibliography{references-cleaned}

% 

\end{document}